\documentclass[
]{iopjournal}

\usepackage{
       graphicx,
       dcolumn,
       bm,
       mathptmx,
       etoolbox,
       hyperref,
       tcolorbox,
       amsmath
}
\usepackage[utf8]{inputenc}
\usepackage[T1]{fontenc}
\hypersetup{colorlinks,linkcolor={blue},citecolor={blue},urlcolor={blue}}

\begin{document}

% \articletype{Paper} %	 e.g. Paper, Letter, Topical Review...

\title{A Fully Electromagnetic Hybrid PIC-Fluid Model for Predictive Fusion Neutron Yield in Dense Plasma Focus}

\author{
       Yinjian Zhao$^1$\orcid{0000-0003-4362-3630}, 
       Zhe Liu$^1$\orcid{0009-0002-3589-9661},
       Qiang Sun$^{2,*}$,
       Qianhong Zhou$^{3,*}$\orcid{0000-0001-8448-7263} and
       Guangrui Sun$^1$
       }

\affil{$^1$School of Energy Science and Engineering, Harbin Institute of Technology, Harbin 150001, People`s Republic of China}

\affil{$^2$School of Electrical Engineering and Automation, Hefei University of Technology, Hefei 230009, People`s Republic of China}

\affil{$^3$Institute of Applied Physics and Computational Mathematics, Beijing 100994, People`s Republic of China}

\affil{$^*$Author to whom any correspondence should be addressed.}

\email{sq19992@mail.ustc.edu.cn, zhou\_qianhong@qq.com}


\begin{abstract}
While magnetic confinement fusion (MCF) and inertial confinement fusion (ICF) remain the primary routes toward controlled fusion, progress is still constrained by energy loss, plasma instabilities, and the cost and complexity of large-scale facilities. The Dense Plasma Focus (DPF) device presents a compact, pulsed-power-driven alternative for producing fusion-relevant conditions and neutron emissions. However, the quantitative prediction of neutron yield in DPF devices poses a significant numerical challenge, primarily due to the imperative of self-consistently resolving kinetic ion behavior, electromagnetic energy coupling, and vacuum field evolution. This complexity often impedes a definitive understanding of the underlying neutron production mechanisms. To address this, we develop a fully electromagnetic hybrid simulation framework: ions are advanced kinetically with particle-in-cell, electrons are a quasi-neutral fluid, and Maxwell’s equations are solved in both plasma and vacuum. The generalized Ohm law includes resistive, electron pressure-gradient, and Hall terms, with a predictor-corrector update for current density. We apply the model to a non-hollow 180 kA DPF geometry similar to the LLNL configuration. The simulated ion density, ion temperature, and axial electric field reproduce sheath formation, axial rundown, radial compression, and post-pinch expansion. The outer sheath front position agrees with fully kinetic benchmarks within 10\% over the available comparison interval. With a compact fit to the D-D fusion cross section, the predicted total neutron yield is $0.296\times10^7$, comparable in order of magnitude to reported fully kinetic results at similar currents and nearly two orders of magnitude higher than earlier hybrid results. This demonstrates that the proposed hybrid framework achieves a favorable balance between computational efficiency and physical fidelity, providing a viable and predictive tool for optimizing DPF devices as compact fusion neutron sources.
\end{abstract}

% \twocolumn

\section{Introduction}
\label{sec:intro}

Nuclear fusion energy, with its advantages of abundant fuel resources, cleanliness, safety, and high energy density, is regarded as an ideal solution for future energy. Current research on controlled nuclear fusion primarily follows two major technological pathways: magnetic confinement fusion (MCF) \cite{zylstra2022burning,matzen2005pulsed} and inertial confinement fusion (ICF) \cite{zohuri2017inertial,brezinsek2021plasma}. However, these large-scale devices commonly face challenges such as massive equipment scale, high construction and operational costs, and limited improvements in energy gain factors, necessitating the exploration of more economical and compact alternative technological approaches. The dense plasma focus device, as a compact pulsed-power device that combines characteristics of both magnetic and inertial confinement\cite{mather1964investigation,filippov1961dense,bernstein1970evidence,schmidt2012fully,lee2014plasma}, has demonstrated unique advantages in nuclear fusion research since its invention in the 1950s. DPF devices can utilize self-generated magnetic fields to efficiently compress plasma, producing high-energy-density plasma in an extremely short time (on the nanosecond scale), offering the potential to achieve fusion conditions\cite{milanese1978evidence, lerner2017confined,lerner2012fusion,cicuttin2015experimental,bernstein1971time,inestrosa2015morphological}. Compared to traditional large-scale fusion devices, DPF devices exhibit notable advantages such as relatively simple structure, lower costs, and strong scalability. Particularly in recent years, with advancements in high-temperature superconducting technology and advanced diagnostic techniques, DPF technology has shown new potential in achieving efficient fusion reactions\cite{lerner2011theory,lerner2023focus,kiai2025double}.

\begin{figure}[htbp]
    \centering
    \includegraphics[width=0.5\linewidth]{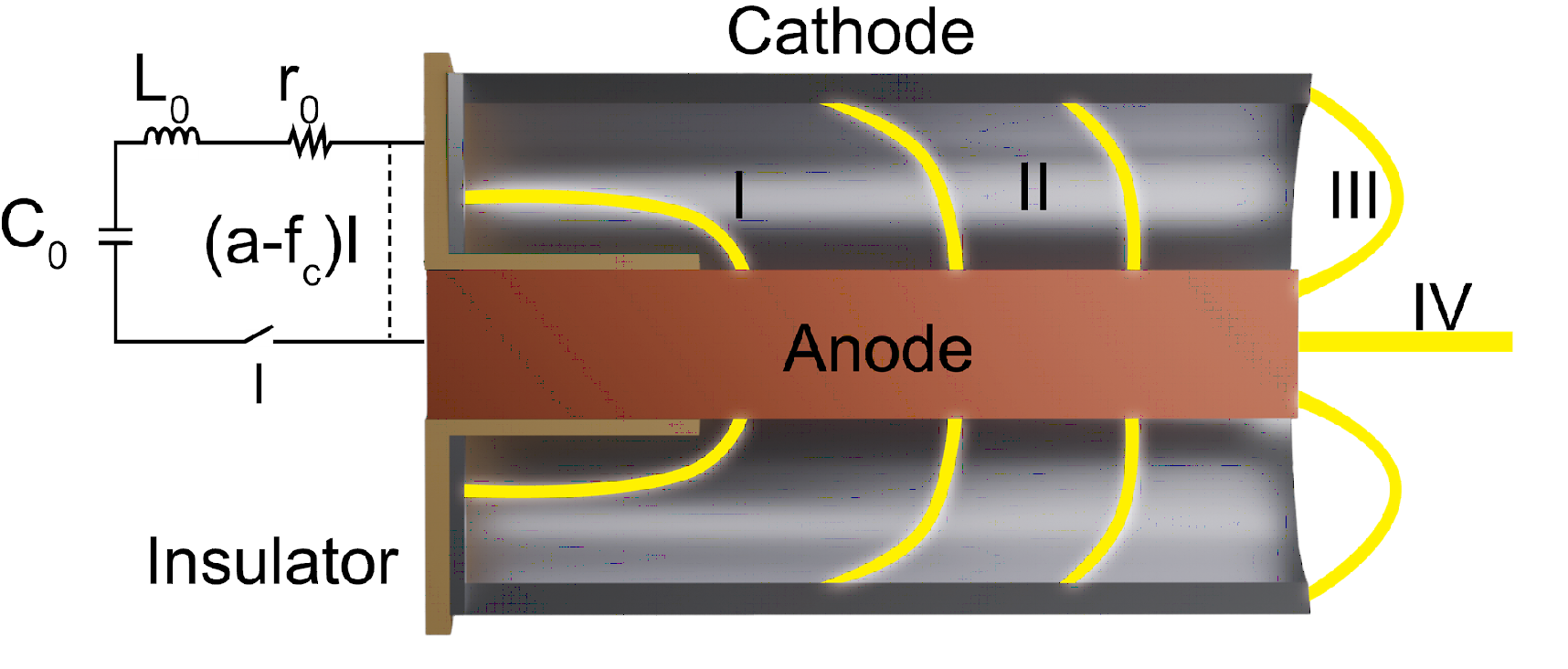}
    \caption{The operational structure of a dense plasma focus device consists of distinct evolution stages, including the flashover stage (I), the rundown stage (II), the run in stage (III), the pinch stage (IV).}
    \label{fig: DPF structure}
\end{figure}

The typical physical evolution of DPF can be divided into four main stages\cite{schmidt2012fully} as Fig. \ref{fig: DPF structure} illustrates:
\begin{enumerate}
   \item  Flashover and liftoff.
   When the high voltage is applied across the coaxial electrodes, the background gas in the area of the insulator is ionized and a current- carrying plasma sheath is created.  The sheath lifts off the insulator surface and closes the circuit, allowing a high current to pass.

   \item  Axial rundown. The plasma sheath is accelerated down the length of the inner electrode by the self-generated azimuthal magnetic field through the \( \mathbf{j} \times \mathbf{B} \) force. During this rundown phase the sheath sweeps up and ionizes neutral gas, acquiring mass and kinetic energy while remaining confined around the anode surface.

   \item Radial run-in. After the sheath reaches the end of the inner electrode, the current channel curves around the anode tip and the Lorentz force develops a substantial radial component. The current sheath then converges toward the axis, forming an imploding plasma column whose radius decreases rapidly and whose inductance increases in time.

   \item  Stagnation and pinch. At the end of the radial implosion the plasma stagnates on axis and forms a dense Z pinch column that is confined by its own azimuthal magnetic field. In this phase the plasma temperature and density reach their maximum values, so that thermal and beam target fusion reactions occur and high energy ions, electrons, X-rays and neutrons are produced.
 \end{enumerate}
 
Strong electromagnetic field evolution, particle acceleration, transport, and energy conversion accompany these steps and collectively constitute the key physical processes addressed in DPF simulations\cite{yousefi2009simulations,bennett2017development,tang2010dense} and experiments\cite{mejia1997some,lerner2011theory,johnson1974study}. These processes directly regulate the formation of energetic-ion populations and, consequently, the neutron yield and pulse characteristics in DPF discharges.

In neutron-source applications of DPF devices, predictive capability for neutron yield is a primary requirement, motivating increasingly physics-rich models. Over the past few decades, a range of modeling and simulation tools have been developed to handle the intricate discharge dynamics of DPF devices. Early studies focused on the canonical coaxial electrode geometry. Foundational experimental configurations and scaling concepts were established by Filippov\cite{filippov1961dense} and Mather\cite{mather1964investigation}. Subsequently, simplified descriptions of current-sheath propagation, energy deposition, and pinch characteristics were provided by semi-empirical models such as the snowplow model and the Lee model\cite{gratton1983two,lee2014plasma}. Although these models offer the benefits of low computational cost and analytical tractability, their ability to capture fine-scale kinetic effects and nonthermal behavior is fundamentally limited. Overall, existing DPF modeling approaches can be grouped into three main classes: semi-empirical snowplow-type models, magnetohydrodynamic (MHD) and fluid descriptions, and fully kinetic particle-in-cell (PIC) simulations.

With increasing computational capabilities, higher-fidelity simulations have been increasingly pursued to improve quantitative prediction of neutron production by resolving pinch formation, instabilities, and energetic-ion generation. A. Schmidt \emph{et al.}\cite{schmidt2012fully,schmidt2014comparisons,schmidt2014fully} employed a fully kinetic approach to carry out high-resolution simulations of DPF discharges, resolving particle dynamics and electromagnetic field evolution during the pinch phase. To improve modeling efficiency, Mathuthu \emph{et al.}\cite{mathuthu2002three} proposed the three-phase model, which splits the discharge process into discrete physical stages\cite{ay2015simulation}. Using the LA-COMPASS code, Li \emph{et al.}\cite{li2017dense} conducted high-fidelity two- and three-dimensional MHD simulations of DPF discharges, providing further insight into pinch formation, related instabilities, and their influence on neutron production. To capture the impact of initial plasma injection profiles and non-ideal resistive effects on the implosion dynamics, Beresnyak \emph{et al.}\cite{beresnyak2018dpf} used a finite-volume MHD scheme to solve the fluid equations for DPF rundown and collapse on a high-impedance generator using the Eulerian 3D Athena code. Jiang \emph{et al.}\cite{jiang2019effect} highlighted how polarity affects ion acceleration and neutron yield by simulating beam and plasma target formation in DPF discharges under various electrode polarities using the fully kinetic CHICAGO PIC code.

These studies have considerably increased our understanding of the underlying physics and enriched DPF modeling from a range of viewpoints. However, they also highlight a persistent challenge for fusion-relevant use cases\cite{lerner2023focus,kiai2025double,gribkov2006summary}: balancing computational cost with the accuracy required for quantitative neutron-yield prediction.

While fully kinetic models can accurately resolve microscopic mechanisms in DPF discharges—such as energy conversion, transport, and nonthermal particle acceleration—their high computational cost makes them unsuitable for neutron-yield optimization, scaling studies, or large-scale parameter scans across operating conditions\cite{schmidt2012fully}. Hybrid PIC-fluid models provide a natural compromise between fully kinetic and fluid descriptions and have been successfully applied in various high-energy-density contexts\cite{thoma2017hybrid}. However, only a few studies have so far touched upon hybrid modeling in the specific context of DPF discharges, mostly in the form of limited kinetic electron components embedded in predominantly MHD or kinetic frameworks\cite{goyon2023comprehensive,goyon2025neutron}. To our knowledge, a systematic ion-PIC/electron-fluid hybrid framework targeted at discharge dynamics and quantitative neutron-yield prediction has not yet been reported.

In particular, fluid models—widely employed in DPF discharge simulations because of their numerical stability and computational efficiency—rely heavily on the continuum approximation. Comparative studies between fluid, hybrid, and fully kinetic simulations of DPF discharges explicitly show that this limits predictive capability for high-energy ions and nonthermal spectral tails by making it difficult to adequately characterize the nonlinear acceleration and collective particle dynamics occurring during the pinch phase\cite{schmidt2012fully}. This limitation is particularly important for neutron-yield prediction, because energetic-ion tails can substantially affect neutron production depending on operating conditions.

Several hybrid PIC-fluid frameworks in high-energy-density physics rely on electromagnetic quasistatic or Darwin-type approximations, which purposefully suppress electromagnetic vacuum modes and light waves to relax time-step constraints and improve efficiency. Although this is beneficial for low-frequency, quasi-static problems, it limits the range of scenarios that may be captured within a single model and makes it difficult to treat electromagnetic radiation and field evolution in extended vacuum regions in a self-consistent manner.

The goal of this work is to establish a hybrid framework that bridges fully kinetic and conventional hybrid descriptions of DPF discharges. The main contributions are twofold. First, in order to self-consistently evolve fields in both the plasma and surrounding vacuum regions, we construct a fully electromagnetic hybrid solver based on a quasi-neutral formulation that couples ion PIC with an electron fluid and solves the full set of Maxwell's equations instead of relying on a Darwin-type approximation. Second, comparison against fully kinetic simulations demonstrates that the proposed hybrid model reproduces key global metrics, in particular neutron yield, with good agreement while substantially reducing computational cost. These findings suggest that an ion PIC/electron-fluid hybrid model can preserve efficiency while achieving near fully kinetic-level accuracy for DPF-relevant observables. Overall, the model not only reproduces the characteristic discharge behavior of DPF devices but also achieves high simulation accuracy at moderate computational cost, providing a physically consistent and computationally feasible tool for investigating high-efficiency energy compression and high-yield DPF operation. Validation on a non-hollow anode configuration shows trends consistent with fully kinetic simulations.

\section{Method}

\subsection{Governing equations}\label{sec:eqs}

The considered governing equations of the DPF system
are as follows.
First, the Faraday's law and the Ampere's law
of the full Maxwell equations are included:
\begin{equation}\label{eq:Faraday}
    \bm{\nabla} \times \bm{E} =
    - \frac{\partial \bm{B}}{\partial t},
\end{equation}
\begin{equation}\label{eq:Ampere}
     \bm{\nabla} \times \bm{B} =
     \mu_0 \left( \bm{J} + \varepsilon_0
     \frac{\partial \bm{E}}{\partial t} \right),
\end{equation}
where $\bm{E}$ is the electric field,
$\bm{B}$ is the magnetic field,
$\bm{J}$ is the current density,
$\mu_0$ is the vacuum permeability,
and $\varepsilon_0$ is the vacuum permittivity.
Then, the generalized Ohm's law is applied:
\begin{equation}\label{eq:Ohm}
    \bm{J} = \sigma \left(\bm{E} + \bm{v}_e \times \bm{B} +
    \frac{\bm{\nabla} p_e}{e n_e} \right),
\end{equation}
where the pressure $p_e = n_e k_B T_e$,
$n_e$ is the electron density,
$T_e$ is the electron temperature,
$\sigma$ is the conductivity,
which is a function of $n_e$ and $T_e$,
and $e$ is the elementary charge.
Next, ions are represented as macro-particles,
and their equations of motion are solved,
\begin{equation}\label{eq:x}
    \frac{d \bm{x}_i}{d t} = \bm{v}_i,
\end{equation}
\begin{equation}\label{eq:v}
   m_i \frac{d \bm{v}_i}{d t} =
   e (\bm{E} + \bm{v}_i \times \bm{B}) - \bm{f},
\end{equation}
where $\bm{x}_i$ and $\bm{v}_i$ denote the ion positions
and velocities,
$\bm{f}$ denotes the ion-electron friction force,
\begin{equation}\label{eq:f}
    \bm{f} = \frac{e \bm{J}}{\sigma}.
\end{equation}
If we substitute Eq.(\ref{eq:Ohm})
into Eq.(\ref{eq:f}),
Eq.(\ref{eq:v}) can also be written as
\begin{equation}\label{eq:v2}
    m_i \frac{d \bm{v}_i}{d t} =
    \frac{1}{n_e}
    \left( \bm{J} \times \bm{B} -
    \bm{\nabla} p_e \right).
\end{equation}
At last, the system is closed by enforcing quasineutrality, $n_e = n_i$, and by assuming ion-electron temperature equilibration, $T_e = T_i$. This single-temperature closure is adopted in all simulations in this work, and its impact on the macroscopic observables and on the neutron yield is assessed a posteriori in Sec.\ref{subsec: para sensitivity}.

\subsection{Solving procedures}\label{sec:procedures}

To better present the numerical method used to solve those governing
equations listed in Sec.\ref{sec:eqs},
we first describe the procedures in the main time loop.
At time step $n$, assume we have known the following
quantities defined at $n$:
the ion positions $\bm{x}_i^n$,
the electric field $\bm{E}^n$;
and the following quantities defined at $n+1/2$:
the ion velocities $\bm{v}_i^{n+1/2}$,
the magnetic field $\bm{B}^{n+1/2}$.

\begin{figure*}[htbp]
    \centering
    \includegraphics[width=0.8\linewidth]{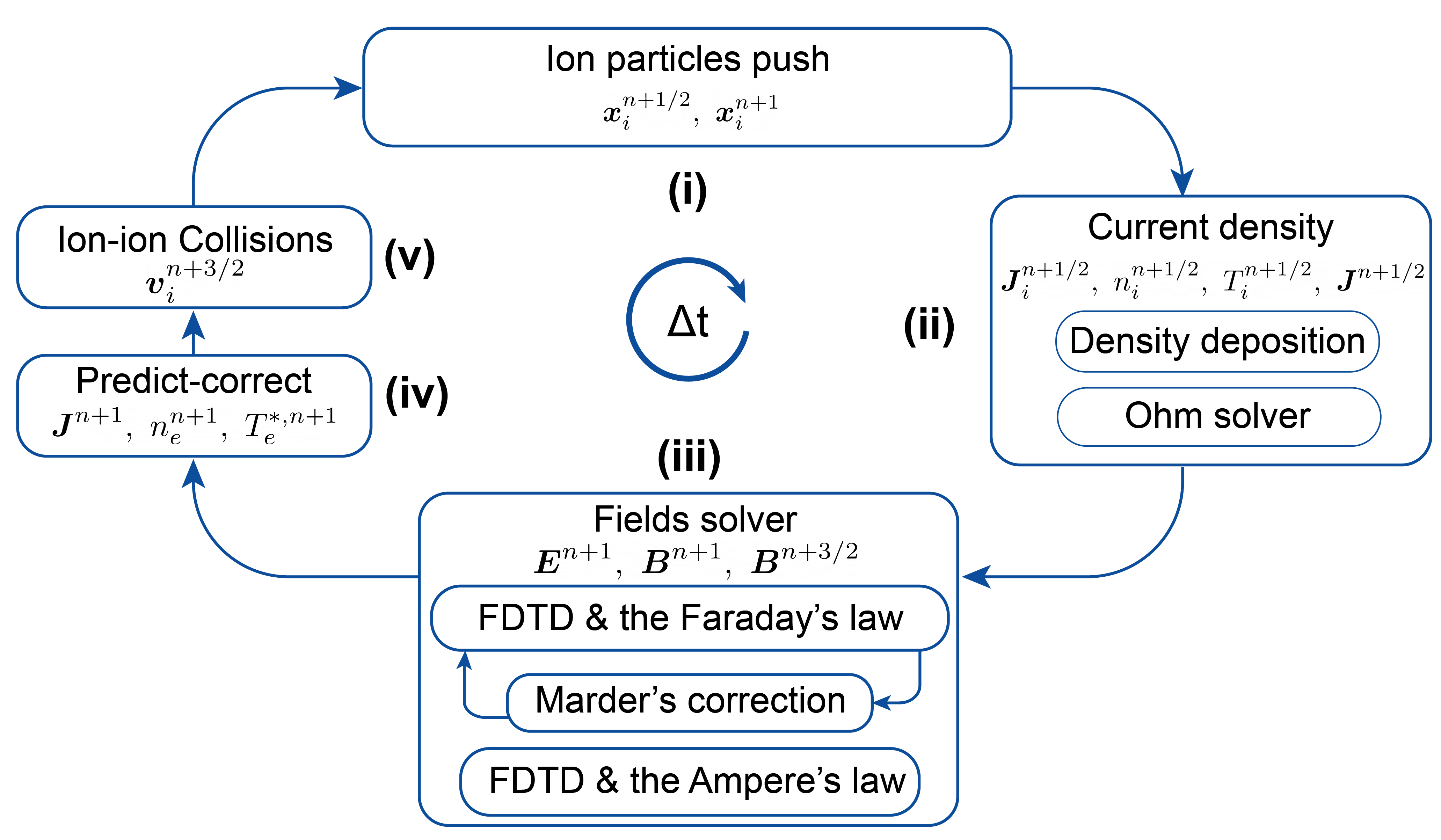}
    \caption{Main time-stepping workflow of the hybrid ion‑PIC/electron‑fluid electromagnetic solver. The loop advances ion macroparticles (Boris), deposits charge/current, solves the Ohm-Ampère closure for the current density, updates fields by FDTD with Marder correction, and applies ion-ion collisions.}
    \label{fig:flowchart}
\end{figure*}

The solving procedures are described below:

(1) Applying the Boris' leap-frog method\cite{Delzanno_2013}
and using $\bm{x}_i^n$, $\bm{v}_i^{n+1/2}$
to solve Eq.(\ref{eq:x}),
we can obtain $\bm{x}_i^{n+1/2}$ by pushing half
time step ($\Delta t/2$),
and $\bm{x}_i^{n+1}$ by pushing a full time step
($\Delta t$).

(2) Applying a cylindrical density deposition
method\cite{Zhao_2022}
and using $\bm{x}_i^n$, $\bm{x}_i^{n+1}$,
we can obtain the ion current density $\bm{J}_i^{n+1/2}$;
using $\bm{x}_i^{n+1/2}$ and $\bm{v}_i^{n+1/2}$,
we can obtain the ion density
$n_i^{n+1/2}$
and temperature
$T_i^{n+1/2}$.

(3) Using $\bm{B}^{n+1/2}$,
$\bm{J}_i^{n+1/2}$,
$n_e^{n+1/2}=n_i^{n+1/2}$,
$T_e^{n+1/2}=T_i^{n+1/2}$,
and $\bm{E}^n$,
the generalized Ohm's law Eq.(\ref{eq:Ohm}) can be solved
along with the Ampere's law Eq.(\ref{eq:Ampere})
to obtain $\bm{J}^{n+1/2}$.
This procedure is described in
Sec.\ref{sec:Ohm} in detail.

(4) Then, the finite difference time domain (FDTD)
method\cite{Inan}
is applied to solve the Ampere's law Eq.(\ref{eq:Ampere}),
using $\bm{B}^{n+1/2}$, $\bm{J}^{n+1/2}$,
and $\bm{E}^n$, to obtain
$\bm{E}^{n+1}$.

(5) Next, the Marder's method\cite{Marder} may be used to
add an additional correction to $\bm{E}^{n+1}$,
such that the continuity equation
can be better preserved,
as described in Sec.\ref{sec:Marder}.

(6) Applying the FDTD method and using
$\bm{E}^{n+1}$ to solve the Faraday's law
Eq.(\ref{eq:Faraday}),
we can half update and full update $\bm{B}^{n+1/2}$ to
obtain $\bm{B}^{n+1}$ and $\bm{B}^{n+3/2}$.

(7) Applying a predictor-corrector method
described in Sec.\ref{sec:PC},
the current density $\bm{J}^{n+1}$ can be obtained,
as well as
$n_e^{n+1}$ computed from $\bm{x}_i^{n+1}$
and
a predictor-corrector temperature
$T_e^{*,n+1}$.

% (7) Now, the velocity equation of motion Eq.(\ref{eq:v})
% can be solved using
% $\bm{J}^{n+1}$, $\bm{B}^{n+1}$, $\bm{E}^{n+1}$,
% and $\bm{v}_i^{n+1/2}$
% to obtain $\bm{v}_i^{n+3/2}$,
% during which
% the first term on the right-hand side
% of Eq.(\ref{eq:v})
% can be solved using the Boris' leap-frog method\cite{Delzanno_2013},
% and the contribution of the
% second $-\bm{f}$ term can be added to the
% velocity separately.

(8) Now, the velocity equation of motion Eq.(\ref{eq:v2})
can be solved using
$\bm{J}^{n+1}$, $\bm{B}^{n+1}$, $n_e^{n+1}$,
and $T_e^{*,n+1}$ to obtain
the updated velocities $\bm{v}_i^{n+3/2}$.
Note that we prefer to use Eq.(\ref{eq:v2})
rather than Eq.(\ref{eq:v}),
because the division of a small $\sigma$ in
Eq.(\ref{eq:f}) may lead to a non-physical huge
force $\bm{f}$.

(9) At last, all ion velocities are
updated via ion-ion collisions
applying Nanbu's Monte Carlo Coulomb collision algorithm
\cite{Nanbu}.

Then,
knowing $\bm{x}_i^{n+1}$,
$\bm{E}^{n+1}$,
$\bm{B}^{n+3/2}$,
and $\bm{v}_i^{n+3/2}$,
the above procedures can be repeated to conduct
the computation at the $n+1$ time step.

\subsection{Ohm solver}\label{sec:Ohm}

With superscripts indicating the time,
Eq.(\ref{eq:Ohm}) can be rewritten as
\begin{eqnarray}\label{eq:Ohm_t}
    \bm{J}^{n+1/2} &=& \sigma^{n+1/2}
    \Bigg(\frac{\bm{E}^{n}+\bm{E}^{n+1}}{2}
    +  \bm{P}^{n+1/2}
    \nonumber \\
    &-& \frac{\bm{J}^{n+1/2}-\bm{J}_i^{n+1/2}}
    {e n_e^{n+1/2}} \times \bm{B}^{n+1/2}
    \Bigg),
\end{eqnarray}
where we use $\bm{P} \equiv \bm{\nabla} p_e/ e n_e$
to represent the pressure gradient term,
$\bm{E}^{n+1/2} \approx (\bm{E}^{n}+\bm{E}^{n+1}) / 2$
to approximate the electric field at $n+1/2$,
and $\bm{J} = \bm{J}_i + \bm{J}_e
= \bm{J}_i - e n_e \bm{v}_e$.
Along with the finite-difference form of the Ampere's law,
\begin{equation}\label{eq:Ampere_t}
    \frac{\bm{E}^{n+1}-\bm{E}^{n}}{\Delta t} =
    c^2 \bm{\nabla} \times \bm{B}^{n+1/2} - \frac{1}{\varepsilon_0}\bm{J}^{n+1/2},
\end{equation}
where $c = (1/\mu_0 \varepsilon_0)^{1/2}$
is the speed of light,
there are two equations with $\bm{J}^{n+1/2}$
and $\bm{E}^{n+1}$ two unknowns,
thus $\bm{J}^{n+1/2}$ can be solved as follows.

Substitute $\bm{E}^{n+1}$ obtained from Eq.(\ref{eq:Ampere_t})
into Eq.(\ref{eq:Ohm_t}),
we can get
\begin{equation}\label{eq:J}
\bm{J}^{n+1/2} (1+\frac{\sigma^{n+1/2} \Delta t}
{2 \varepsilon_0}) =
\bm{A} - \frac{\sigma^{n+1/2}}{e n_e^{n+1/2}}
\bm{J}^{n+1/2} \times \bm{B}^{n+1/2},
\end{equation}
where $\bm{A}$ is a known vector
\begin{equation}\label{eq:A}
\frac{\bm{A}}{\sigma^{n+1/2}} \equiv  
\bm{E}^n + \Bigg( \frac{c^2 \Delta t}{2} \bm{\nabla} +
\frac{\bm{J}_i^{n+1/2}}{e n_e^{n+1/2}} \Bigg)
\times \bm{B}^{n+1/2}
+ \bm{P}^{n+1/2}.
\end{equation}

In cylindrical 2D-RZ geometry,
\begin{equation}
    \bm{J} \times \bm{B} =
    -\hat{r} J_z B_\theta + \hat{z} J_r B_\theta,
\end{equation}
where $\hat{r}$ and $\hat{z}$
denote the radial and axial direction unit vectors,
subscripts $r$, $\theta$, and $z$ denote
the radial, azimuthal, and axial components, respectively.
After some algebra,
we can obtain
\begin{eqnarray}\label{eq:JrJz}
    J^{n+1/2}_r &=& \frac{D A_r + C A_z}{D^2+C^2},
    \nonumber \\
    J^{n+1/2}_z &=& \frac{D A_z - C A_r}{D^2+C^2},
\end{eqnarray}
where
$A_r$ and $A_z$ are the radial and axial components
of the vector $\bm{A}$,
\begin{equation}
    C \equiv \frac{\sigma^{n+1/2}}
     {e n_e^{n+1/2}} B^{n+1/2}_\theta,
\end{equation}
and
\begin{equation}
    D \equiv 1 + \frac{\sigma^{n+1/2} \Delta t}
    {2 \varepsilon_0}.
\end{equation}
Therefore, Eq.(\ref{eq:JrJz}) is eventually used
to obtain $\bm{J}^{n+1/2}$.

If we further ignore the Hall term,
i.e., the $-\bm{J}^{n+1/2} \times
\bm{B}^{n+1/2} / (e n_e^{n+1/2})$
term in Eq.(\ref{eq:Ohm_t}),
the computations in Eq.(\ref{eq:JrJz})
reduce to
\begin{equation}\label{eq:JrJznoHall}
    \bm{J}^{n+1/2} = \bm{A}/D.
\end{equation}

\subsection{Marder's correction}\label{sec:Marder}

The density deposition method
applied\cite{Zhao_2022}
preserves the continuity equation rigorously,
but only when both electrons and ions are represented by
macro-particles.
Because in this hybrid PIC algorithm,
electrons are represented by fluid,
the continuity equation may not be conserved,
thus a correction may be needed to make sure the
simulation is stable and correct.
We therefore apply Marder's correction\cite{Marder}
after the procedure (4) in Sec.\ref{sec:procedures}
to correct the electric field as follows:
\begin{equation}
    \bm{E}^{n+1} \dot{=} \bm{E}^{n+1} +
    d \bm{\nabla} \left(\bm{\nabla} \cdot \bm{E}^{n+1} -
    \frac{\rho}{\varepsilon_0} \right),
\end{equation}
where
$\dot{=}$ denotes the assignment operation,
$\rho$ is the charge density and is zero
due to the quasi-neutral assumption $n_e=n_i$,
$d$ is the Marder factor,
which is chosen small enough not to affect
adversely the stability
but large enough to perform the desired function.

\subsection{Predictor-corrector method}\label{sec:PC}

Because $\bm{J}^{n+1}$ is needed to solve
the ion equation of motion Eq.(\ref{eq:v}),
but only $\bm{J}^{n+1/2}$ is known after procedure (3).
We thus need a
predictor-corrector method to obtain a good estimation
of $\bm{J}^{n+1}$ to make the simulation more stable.
The procedures of the predictor-corrector method
proposed is described as
follows: 

(a) Apply a time-average linear approximation to obtain
\begin{equation}
    \bm{J}^{*,n+1} = 2 \bm{J}^{n+1/2} - \bm{J}^{n},
\end{equation}
where $\bm{J}^{n}$ is stored from the previous time step,
and $\bm{J}^{0} = \bm{J}^{1/2}$ can be set for the first
time step.

(b) Using $\bm{J}^{*,n+1}$ to replace $\bm{J}^{n+1}$
and applying the
procedure (7) in Sec.\ref{sec:procedures},
we can obtain estimated ion velocities
$\bm{v}_i^{*,n+3/2}$.

(c) Apply another time-average linear approximation
to obtain the ion velocities
\begin{equation}
    \bm{v}_i^{*,n+1} =
    \frac{\bm{v}_i^{*,n+3/2}+\bm{v}_i^{n+1/2}}{2},
\end{equation}
thus we can compute $\bm{J}_i^{*,n+1}$,
$T_i^{*,n+1}$ as well as $n_i^{n+1}$ obtained
from $\bm{x}_i^{n+1}$.

(d) Then, we can directly solve the generalized Ohm's law
to obtain the predicted current density
without using Ampere's law,
since $\bm{E}^{n+1}$ has been obtained already in the procedure (4)
in Sec.\ref{sec:procedures},
\begin{eqnarray}\label{eq:Ohm_t2}
    \bm{J}^{n+1} &=& \sigma^{*,n+1}
    \Bigg(\bm{E}^{n+1}
    +  \bm{P}^{*,n+1}
    \nonumber \\
    &-& \frac{\bm{J}^{n+1}-\bm{J}_i^{*,n+1}}
    {e n_e^{n+1}} \times \bm{B}^{n+1}
    \Bigg),
\end{eqnarray}
where $\sigma^{*,n+1}$ is now a function of
$T_e^{*,n+1}=T_i^{*,n+1}$
and $n_e^{n+1} = n_i^{n+1}$,
$\bm{P}^{*,n+1} = \bm{\nabla} (n_e^{n+1} k_B T_e^{*,n+1})
/ e n_e^{n+1}$.
Eq.(\ref{eq:Ohm_t2}) can be solved similarly as
the way introduced in Sec.\ref{sec:Ohm}.
After some algebra, we can obtain
\begin{eqnarray}\label{eq:JrJzc}
    J_r^{n+1} &=& \frac{
    A_r^* + C^* A_z^* }
    {1 + (C^*)^2} \nonumber \\
    J_z^{n+1} &=& \frac{
    A_z^* - C^* A_r^* }
    {1 + (C^*)^2}
\end{eqnarray}
% \begin{eqnarray}
%     J_r^{n+1} &=& \frac{1}{1 + (C^*)^2}
%     \Bigg[
%     \sigma^{*,n+1} (
%     E_r^{n+1} + C^* E_z^{n+1} ) \nonumber \\
%     &-&
%     C^* J_{i,z}^{*,n+1} +
%     (C^*)^2 J_{i,r}^{*,n+1}
%     \Bigg], \nonumber \\
%     J_z^{n+1} &=& \frac{1}{1 + (C^*)^2}
%     \Bigg[
%     \sigma^{*,n+1} (
%     E_z^{n+1} - C^* E_r^{n+1} ) \nonumber \\
%     &+&
%     C^* J_{i,r}^{*,n+1} +
%     (C^*)^2 J_{i,z}^{*,n+1}
%     \Bigg],
% \end{eqnarray}
where $\bm{A}^*$ is a known vector
\begin{equation}
    \frac{\bm{A}^*}{\sigma^{*,n+1}} \equiv  
    \bm{E}^{n+1} + \bm{P}^{*,n+1} +
    \frac{1}{e n_e^{n+1}}
    \bm{J}_i^{*,n+1} \times \bm{B}^{n+1},
\end{equation}
and $C^*$ is defined as
\begin{equation}
    C^* \equiv \frac{\sigma^{*,n+1}}{e n_e^{n+1}}
    B_\theta^{n+1}.
\end{equation}
In addition, if we ignore the Hall term,
Eq.(\ref{eq:JrJzc}) can be reduced to
\begin{equation}
    \bm{J}^{n+1} = \bm{A}^*.
\end{equation}

\begin{figure*}[htbp]
    \centering
    \includegraphics[width=0.9\linewidth]{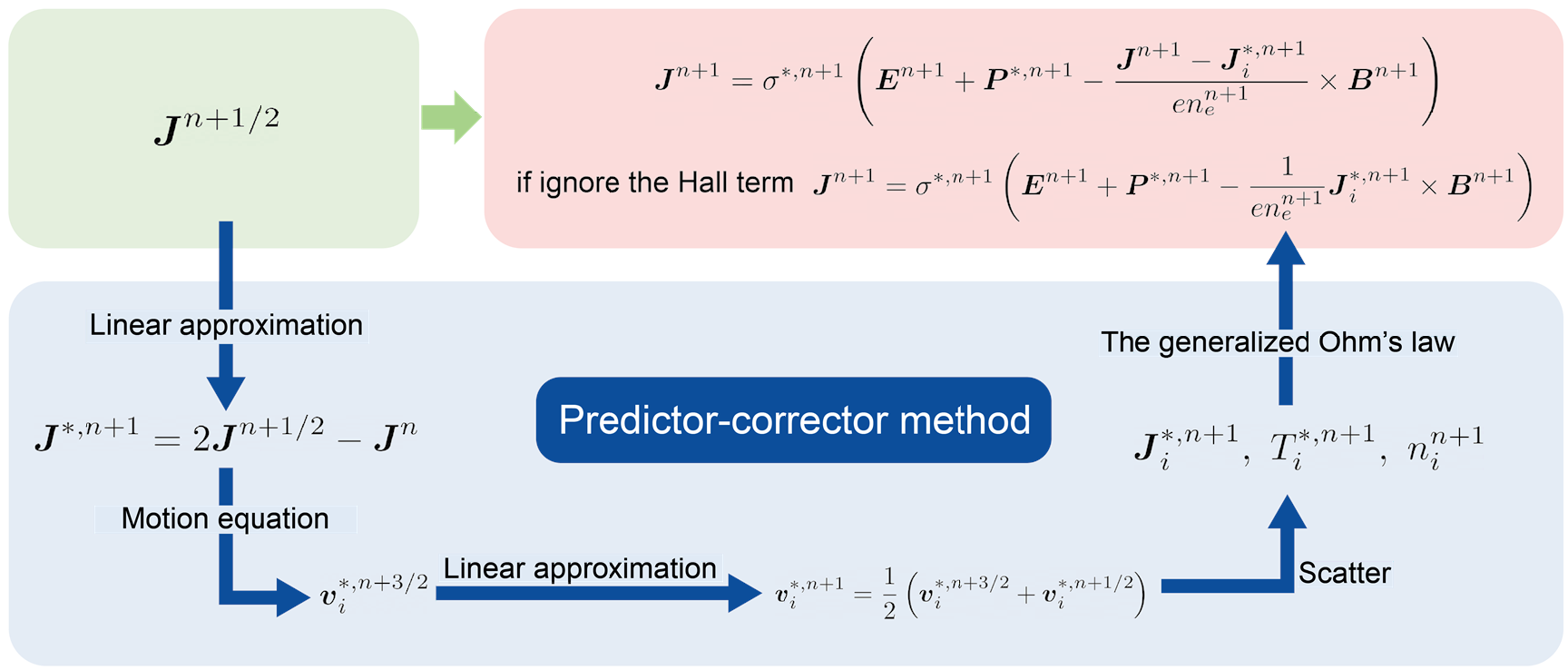}
    \caption{Predictor-corrector scheme used to estimate the end‑of‑step current density. The method time‑extrapolates $J$, performs a provisional ion update, rebuilds $n_e$, $T_e$ and $J_i$, and then solves the generalized Ohm’s law to obtain a stable $J^{n+1}$.}
    \label{fig:flowchart of predict correct}
\end{figure*}

In later sections,
% the performance of this predictor-corrector method
% is compared with two approximations using
% $\bm{J}^{n+1} \approx \bm{J}^{n+1/2}$
% and $\bm{J}^{n+1} \approx \bm{J}^{*,n+1}$,
it is found that this predictor-corrector method
is necessary to obtain more accurate simulation results
in the long run.

\subsection{Conductivity treatment}\label{sec:sigma}

Throughout the simulation, the conductivity
is treated as follows.
According to the local density and temperature,
we compute
\begin{equation}
    \sigma_0 = \frac{e^2 n_e}{m_e \nu_{ei}},
\end{equation}
where $\nu_{ei}$ is the electron-ion collision frequency,
\begin{equation}
    \nu_{ei} = \frac{n_e \ln{\Lambda}}{3.5 \times 10^{11}
    T_e^{3/2}},
\end{equation}
$T_e$ is in electron volts,
$\ln{\Lambda}$ is the Coulomb logarithm, where

\begin{equation}
    \Lambda = n_{\rm e}\lambda^3_D,\quad \lambda_D = \sqrt{\frac{\epsilon_0 k_{\rm B}T_{\rm e}}{n_{\rm e}e^2}}.
\end{equation}

We set $\sigma$ according to
\begin{eqnarray}
    \sigma = 
    \begin{cases}
        0,                        & \text{if } n_e/n_0 \in [0,0.1),\\[4pt]
        (n_e/n_0)^3 \,\sigma_0,   & \text{if } n_e/n_0 \in [0.1,1),\\[4pt]
        \sigma_0,                 & \text{if } n_e/n_0 \in [1,\infty),
    \end{cases}
\end{eqnarray}
where $n_0$ denotes the initial background plasma density, so that $\sigma$ transits smoothly from $\sigma_0$ in the plasma to $0$ in the vacuum.

To avoid excessively large values of $\sigma$ that would make the explicit field update numerically stiff, we apply an additional CFL type limiter. Neglecting spatial derivatives, Amp\`ere's law with an Ohmic current,
\begin{equation}
    \varepsilon_0 \frac{\partial \mathbf{E}}{\partial t} = -\sigma \mathbf{E},
\end{equation}

implies an Ohmic relaxation time
\begin{equation}
  \tau_\sigma = \frac{\varepsilon_0}{\sigma}.
\end{equation}
We require that this relaxation time is not resolved by fewer than $C_\sigma$ time steps,
\begin{equation}
  \tau_\sigma \ge C_\sigma \Delta t
  \qquad\Rightarrow\qquad
  \sigma \le \sigma_{\mathrm{CFL}} \equiv \frac{C_\sigma \varepsilon_0}{\Delta t},
\end{equation}
where $\Delta t$ is the global time step and $C_\sigma$ is a safety factor of order unity. The effective conductivity used in the Maxwell update is then taken as
\begin{equation}
  \sigma_{\mathrm{eff}} = \min\!\bigl[\sigma(n_e),\,\sigma_{\mathrm{CFL}}\bigr].
\end{equation}
In practice this limiter prevents the Ohmic term from violating the stability constraint of the explicit FDTD scheme, while leaving the conductivity in the dense plasma region essentially unchanged.

\subsection{Grid system and the simulation domain}
\label{sec:grid}

\begin{figure}[htbp]
  \centering
  \includegraphics[width=0.5\linewidth]{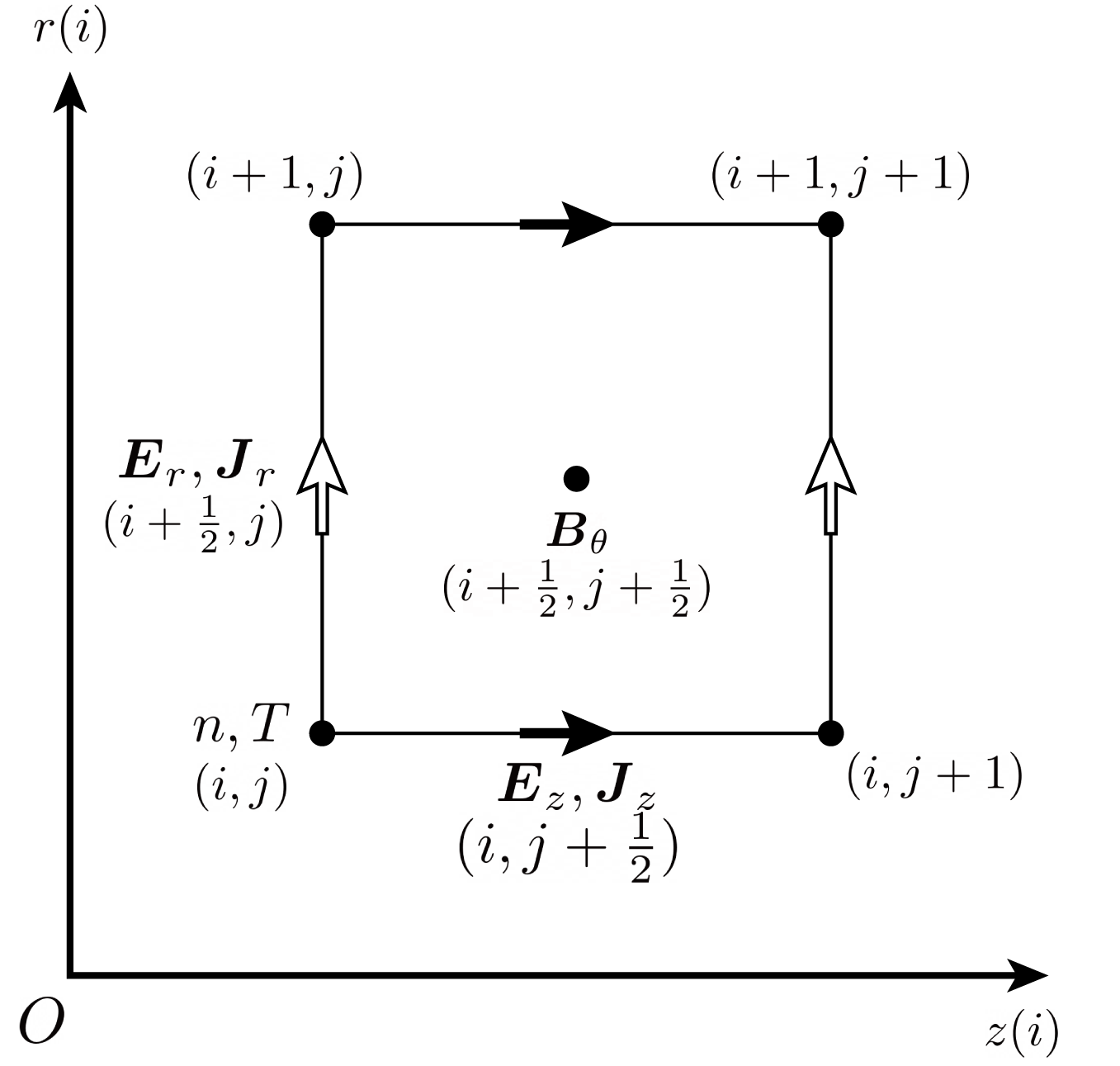}
  \caption{Staggered Yee grid in 2D‑RZ. Densities/temperatures $(n,T)$ are defined at grid nodes; $B_\theta$ at cell centers; $(E_r,E_z)$ and $(J_r,J_z)$ at face midpoints.}
  \label{fig:grid}
\end{figure}

Since the FDTD method is used to solve the
Maxwell equations,
the staggered Yee's grid is applied,
as illustrated in Fig.\ref{fig:grid}.
The plasma density and temperature quantities
are defined at the grid points;
the magnetic field is defined
at the cell center;
the electric field and the current density
quantities are defined at
the midpoints of cell sides.

\begin{figure}[htbp]
  \centering
  \includegraphics[width=0.5\linewidth]{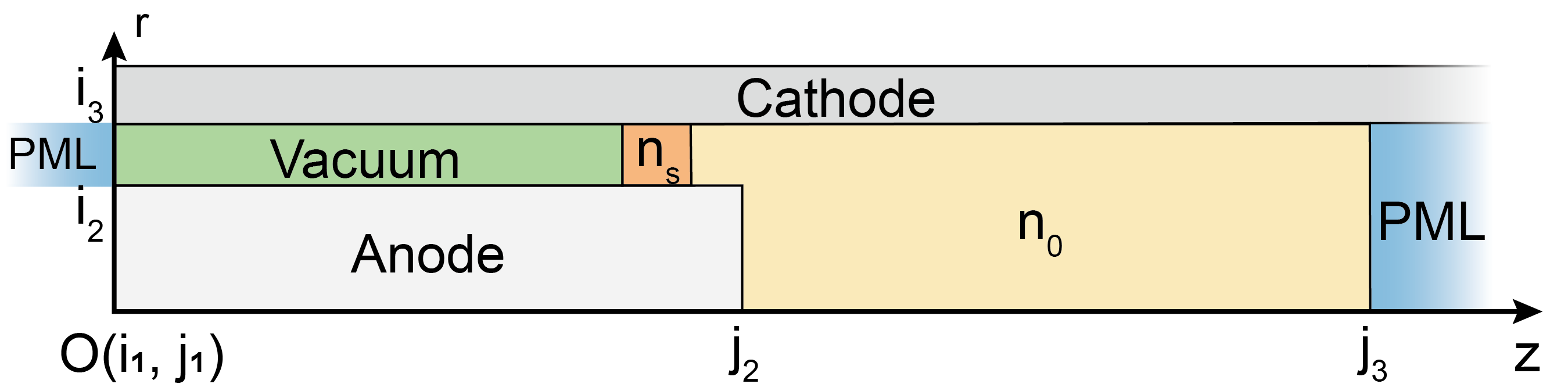}
  \caption{Computational domain and boundary conditions for the LLNL‑like DPF geometry. The left and the right boundary employs an axial PML; the upper boundary and shaded regions are perfect conductors. }
  \label{fig:domain}
\end{figure}

The simulation domain is set up as illustrated
in Fig.\ref{fig:domain},
in which
the sheath plasma with an initial density $n_s$
and
the background plasma with an initial density
$n_0$ are placed accordingly.
The left boundary at $z=0$ (or $j=0$)
and $i\in[i_2,i_3]$ is open;
as well as the right boundary at $j=j_3$,
where the Perfectly Matched Layer
(PML)
absorbing boundary condition
\cite{Inan}
is applied for
the outgoing electromagnetic waves.
Besides the conductor labeled
in Fig.\ref{fig:domain},
the top boundary $i=i_3$ is also treated as a conductor,
such that electromagnetic waves are reflected.
Particles entering either the conductor region
or the PML region
are absorbed and deleted.

\section{Simulation setup}
\label{subsec: simulation setup}

\subsection{Grid and particles}
\label{subsec: grid and particles}

For the implosion to pinch phase in DPF, this paper establishes a particle PIC-electron fluid model based on the quasi-neutral assumption and develops hybrid simulation code. The ion PIC-electron fluid hybrid code consists of three fundamental modules: the ion PIC module, electron fluid module, and electromagnetic field module. The PIC module solves ion velocities and positions, the fluid module calculates electron velocities, temperature, and density, while the electromagnetic field module computes global electric and magnetic fields. Through communication modules and optimized hybrid simulation iteration processes, self-consistent coupling of ions, electrons, and electromagnetic fields (i.e., particle-fluid hybrid simulation) is achieved.

The electrode geometry aligns with LLNL's hundred-kA compact DPF experimental setup. A 5-cm-long anode serves as the inner electrode with radius 1.0 cm. The cathode measures 10 cm in length. According to the CFL requirement in Eq.\ref{eq: CFL}, the CFL factor ($C$) in the simulation is 0.9, and $\Delta r$ and $\Delta z$ are both 0.02 cm, hence the time step of the simulation is $\Delta t = 4.25\times 10^{-13}$ s. Simulation parameters are listed in Tab.\ref{tab:sim-params}.

\begin{equation}
    \Delta t = C \cdot \left( \frac{1}{c\sqrt{ \frac{1}{\left(\Delta r\right)^2} +  \frac{1}{\left(\Delta z\right)^2}}} \right)
    \label{eq: CFL}
\end{equation}

\begin{table}[htbp]
\centering
\caption{Key simulation parameters for the DPF implosion-to-pinch hybrid PIC-electron fluid model.}
\label{tab:sim-params}
\begin{tabular}{ll}
\hline
\textbf{Parameter} & \textbf{Value} \\
\hline
% \multicolumn{2}{l}{\textbf{Geometry}} \\
Anode length $L_a$ & 5 cm \\
Anode radius $r_a$ & 1.0 cm \\
Cathode length $L_c$ & 10 cm \\
% \\[-8pt]
% \multicolumn{2}{l}{\textbf{Domain}} \\
Coordinate system & 2D axisymmetric $(r,z)$ \\
Physical domain size & $1.5$ cm $\times$ $10$ cm \\
% \\[-8pt]
% \multicolumn{2}{l}{\textbf{Grid}} \\
Radial cells $N_r$ & 77 \\
Axial cells $N_z$ & 522 \\
Cell size $(\Delta r,\Delta z)$ & $(0.02,\,0.02)$ cm \\
% \\[-8pt]
% \multicolumn{2}{l}{\textbf{PML}} \\
PML absorbing layers (axial) & 20 \\
% \\[-8pt]
% \multicolumn{2}{l}{\textbf{Time}} \\
Time step $\Delta t$ & $4.25\times 10^{-13}$ s \\
% \\[-8pt]
% \multicolumn{2}{l}{\textbf{Plasma (initial)}} \\
Sheath thickness & $\sim 1$ mm \\
Sheath number density $n_{s,0}$ & $3.3\times 10^{23}$ m$^{-3}$ \\
Background number density $n_{n,0}$ & $6.7\times 10^{22}$ m$^{-3}$ \\
% \\[-8pt]
% \multicolumn{2}{l}{\textbf{Particles}} \\
Particles in background region & $500,000$ \\
Particles in sheath region & $26,060$ \\
% \\[-4pt]
% \multicolumn{2}{l}{\textbf{Start}} \\
% Initial phase & End of rundown \\

\hline
\end{tabular}

\end{table}

\begin{figure}[htbp]
    \centering
    \includegraphics[width=0.5\linewidth]{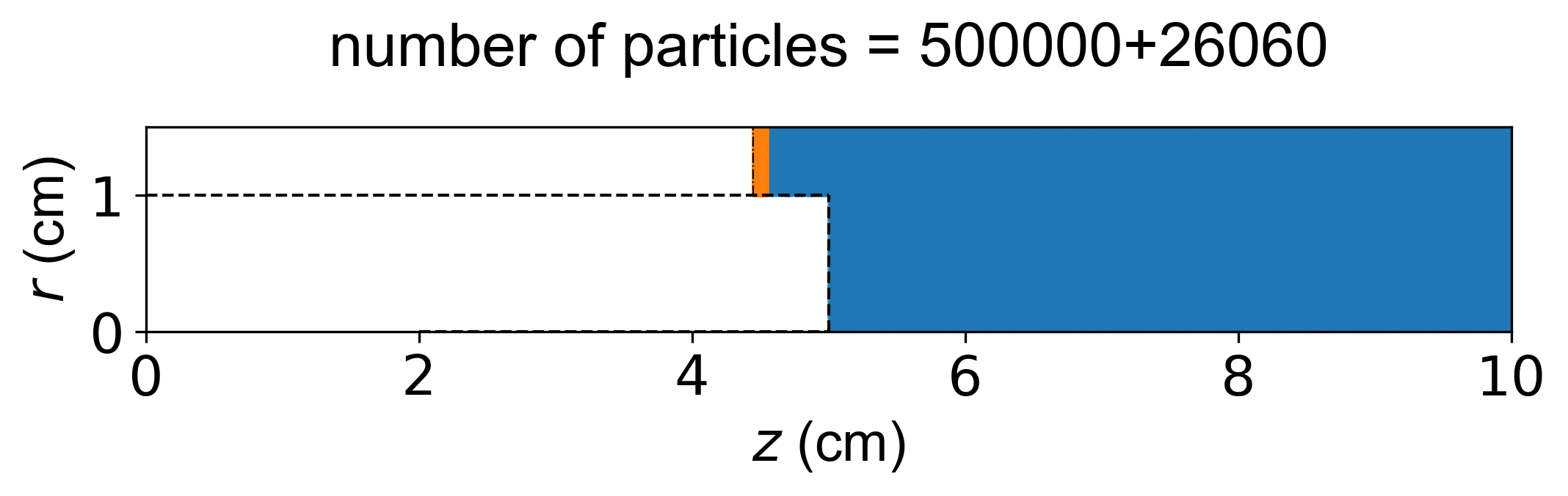}
    \caption{ Initial conditions at the end of rundown. A $\sim$1\,mm plasma sheath (orange) centered near $z\!\approx\!4.5$\,cm with $n_{s,0}=3.3\times10^{23}\,\mathrm{m^{-3}}$ is followed upstream by deuterium plasma number density of $n_{n,0}=6.7\times10^{22}\,\mathrm{m^{-3}}$ and by vacuum downstream.}
    \label{fig: initial conditions}
\end{figure}

The ion macroparticles are initialized by Monte Carlo sampling in cylindrical
coordinates. A first population of $N_{1}=5\times10^{5}$ macroparticles
represents the pre filled deuterium background with number density
$n_{0}=6.7\times10^{22}\ \mathrm{m}^{-3}$. Particles are distributed uniformly
in the volume $0\le r\le1.5\ \mathrm{cm}$, $5\ \mathrm{cm}\le z\le10\ \mathrm{cm}$,
excluding the pre ionized sheath region and the electrode metal by rejecting
samples that fall inside those volumes (see Fig.\ref{fig: initial conditions}). The background velocities
are drawn from isotropic Maxwellian distributions with temperature
$T_{1}\simeq0.026\ \mathrm{eV}$ ($\approx300\ \mathrm{K}$) for both ions and
electrons, so that the background plasma is initially cold and stationary apart
from thermal fluctuations.

A second population of $N_{2}=2.606\times10^{4}$ macroparticles represents the
pre accelerated current sheath at the end of rundown. These ions are placed
uniformly in a thin slab of axial thickness $\delta z\simeq1\ \mathrm{mm}$
adjacent to the anode surface, spanning $1.0\ \mathrm{cm}\le r\le1.5\ \mathrm{cm}$
and $4.45\ \mathrm{cm}\le z\le4.55\ \mathrm{cm}$, as indicated by the orange
region in Fig.\ref{fig: initial conditions}. Their velocities are sampled from drifting Maxwellian
distributions with temperature $T_{2}\simeq7.2\times10^{5}\ \mathrm{K}$
($\approx62\ \mathrm{eV}$). The radial and azimuthal components have zero mean,
whereas the axial component has an additional drift
$v_{d}\simeq1.1\times10^{5}\ \mathrm{m}\,\mathrm{s}^{-1}$ along $+z$. This
procedure produces a narrow, high density sheath that already carries the axial
current at $t=0$ and approximates the end of rundown state seen in fully kinetic
simulations. The electron fluid is initialized consistently by setting
$n_{e}=n_{i}$ and $T_{e}=T_{i}$ in each cell.

A 2D axisymmetric coordinate system is employed. The r-direction contains 77 grids (75 for main DPF region and 2 boundary grids) and the z-direction 522 grids (500 for 10-cm main region, 20 PML absorbing layers and 2 boundary grids). This configuration models a 1.5 cm $\times$ 10 cm region with additional boundary control grids.

The simulation initiates at the rundown phase conclusion. Fig.\ref{fig: initial conditions} displays initial macroparticle distribution, requiring $\ge 10$ particles per cell in the about 1-mm plasma sheath (orange regions). The initial number density of the sheath is 3.3$\times 10^{23}$ m$^{-3}$, with velocity combining matter and radiation at extreme thermal and drift components. The background deuterium plasma number density is 6.7$\times 10^{22}$ m$^{-3}$.

The computational domain extends to a maximum radius of $R_{\max} = 1.56~\mathrm{cm}$ and an axial length of $Z_{\max} = 10.46~\mathrm{cm}$, and is discretized on a uniform cylindrical grid with $\Delta r = \Delta z = 0.02~\mathrm{cm}$ ($78 \times 523$ cells). The pre-filled deuterium column occupies $r \le 1.5~\mathrm{cm}$ and $5~\mathrm{cm} \le z \le 10~\mathrm{cm}$, corresponding to $75 \times 250$ cells. From the simulated ion-density profiles, the axial thickness of the current sheath is of the order $\delta_{\text{sheath}} \approx 0.15\text{ to }0.20~\mathrm{cm}$, which is resolved by about $8$ to $10$ cells in the $z$ direction, while the pinch radius is $r_p \approx 0.5~\mathrm{cm}$, corresponding to roughly $25$ cells in the radial direction. The present resolution is therefore sufficient to capture the macroscopic sheath motion and pinch structure.

The electromagnetic fields are advanced with an explicit FDTD solver. For $\Delta r = \Delta z$ the electromagnetic CFL condition $c\,\Delta t \le \Delta r / \sqrt{2}$ yields a stability limit of $\Delta t_{\text{CFL}} \approx 4.7 \times 10^{-13}~\mathrm{s}$ for the present grid. In all simulations the time step is chosen such that $\Delta t \le \Delta t_{\text{CFL}}$ and $\Delta t \ll \Delta r / v_{i,\text{max}}$, where $v_{i,\text{max}}$ is the characteristic ion drift speed, so that both wave propagation and ion dynamics are adequately resolved.

Initially, the pre-filled plasma region ($r \le 1.5~\mathrm{cm}$, $5~\mathrm{cm} \le z \le 10~\mathrm{cm}$) is sampled with $5 \times 10^{5}$ ion macroparticles distributed over $75 \times 250$ cells, giving an average of about $27$ particles per cell. An additional $2.6 \times 10^{4}$ macroparticles are loaded in the pre-ionized layer near the anode. These values lie within the commonly adopted range of tens of particles per cell in PIC simulations of dense plasmas, and in the present runs the current density and field distributions exhibit only modest statistical noise.

\subsection{Circuit and magnetic field}
\label{subsec: circuit and B}

The simulation requires solving current $I$ from the external circuit, then calculating magnetic boundary conditions at the injection port based on the current at each timestep. The azimuthal magnetic field in the simulation is computed via Eq.\ref{eq:B theta}.

\begin{equation}
    B_\theta = \frac{\mu I}{2\pi r}
    \label{eq:B theta}
\end{equation}

The current equation for the external circuit model is:
\begin{equation}
\frac{\mathrm{d}\left(L_0 I\right)}{\mathrm{d}\, t} =  V_0 - r_0I - U_{\text{DPF}}  - \frac{1}{C_0}\int I \mathrm{d}\,t
\label{eq: U DPF}
\end{equation}

where $L_0=110$ $nH$ represents circuit inductance, $V_0=10$ $kV$ circuit voltage, $r_0=12$ $m\Omega$ stray resistance, and $C_0=20$ $\mu F$ circuit capacitance \textemdash all predefined parameters. The DPF system voltage $U_{\text{DPF}}$ must be calculated through Eq.\ref{eq: U DPF}, involving magnetic field integration over the entire DPF system followed by time differentiation to obtain the DPF voltage at each timestep.

\begin{equation}
    U_{\rm{DPF}} = \frac{{\rm{d}}\,\left( \int B{\rm{d}}\,s \right)}{{\rm{d}}\,t}
\end{equation}

This equation is an ordinary differential equation and is discretized using the following first‑order explicit scheme:
\begin{equation}
   \frac{I^{n+1}-I^{n}}{\Delta t} = \frac{1}{L_0}\!\left( V_0 - r_0 I^{n} - \frac{Q^{n}}{C_0} - U_{\mathrm{DPF}}^{\,n} \right)
   \label{eq: I}
\end{equation}

\begin{equation}
    \frac{Q^{n+1}-Q^{n}}{\Delta t} = I^{n}
    \label{eq: QI}
\end{equation}
Here, $Q=\int I\,\mathrm{d}t$ denotes the circuit charge.
The initial conditions are $I^{0}=1.773\times10^{4}\ \mathrm{A}$ and
$Q^{0}=0.218\ \mathrm{C}$.
The circuit parameters are
$L_0=1.1\times10^{-7}\ \mathrm{H}$,
$V_0=1.5\times10^{4}\ \mathrm{V}$,
$r_0=1.2\times10^{-2}\ \Omega$, and
$C_0=2.0\times10^{-5}\ \mathrm{F}$.
This update yields the current $I$ at each time step, from which the
magnetic‑field boundary condition at the injection port is obtained.

In Eq.\ref{eq: U DPF}-\ref{eq: QI} we use the same lumped $RLC$ circuit model to describe the
generator that drives the DPF, but two different values associated with the
parameter $V_{0}$ appear. The set
$L_{0}=110\ \mathrm{nH}$, $r_{0}=12\ \mathrm{m\Omega}$,
$C_{0}=20\ \mu\mathrm{F}$, and $V_{0}=10\ \mathrm{kV}$
characterizes the nominal charging of the capacitor bank and is used to define
the continuous circuit equation in Eq.\ref{eq: U DPF}. In the discrete update
\ref{eq: I}-\ref{eq: QI} we instead specify the instantaneous state of the circuit at the
beginning of the hybrid simulation, which corresponds to a later time in the
discharge close to the end of rundown. The initial values $(I^{0},Q^{0})$ and
the effective voltage $V_{0}=1.5\times10^{4}\ \mathrm{V}$ are chosen such that
the integrated current waveform reproduces the $\sim 180\ \mathrm{kA}$ plateau
required to drive the sheath motion shown in Fig.\ref{fig: iu}. In other words, the
lower value of $V_{0}$ refers to the nominal bank charging voltage, whereas the
higher value represents the capacitor voltage at the instant when the hybrid
implosion-to-pinch simulation is initialized.

\section{Simulation results and discussion}
\label{sec:simulation}
% \subsection{Baseline}
% \subsection{Baseline case}
\subsection{Simulation results for DPF discharge}
\label{subsec: result}

\begin{figure}[htbp]
    \centering
    \includegraphics[width=0.5\linewidth]{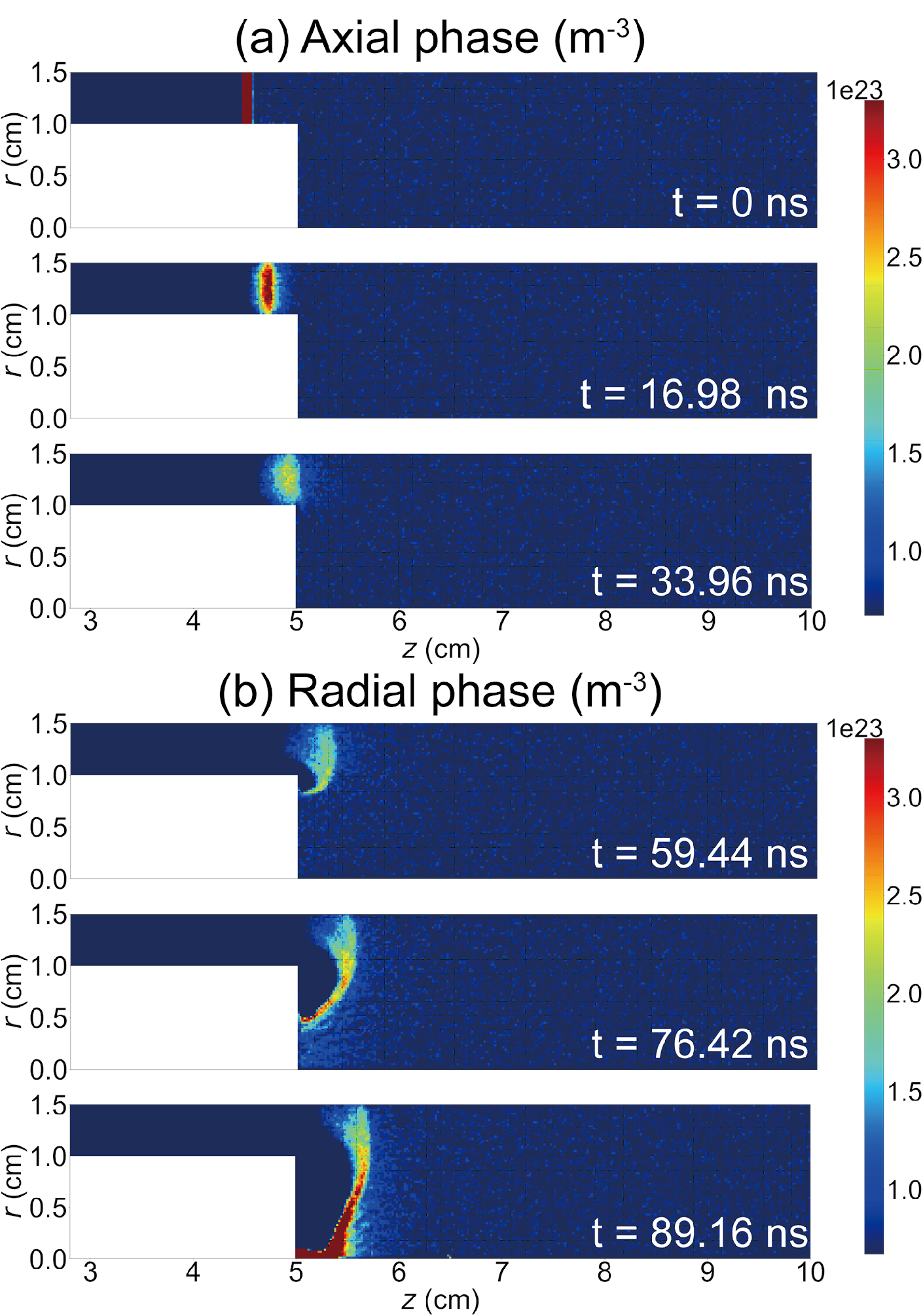}
    \caption{Time evolution of ion number density from the hybrid simulation. The axial phase (top panels) shows sheath propagation along the anode surface, and the radial phase (bottom panels) shows the sheath wrapping around the anode tip and imploding toward the axis to form a dense pinch column.}
    \label{fig: sgm 001}
\end{figure}

\begin{figure}[htbp]
    \centering
    \includegraphics[width=0.5\linewidth]{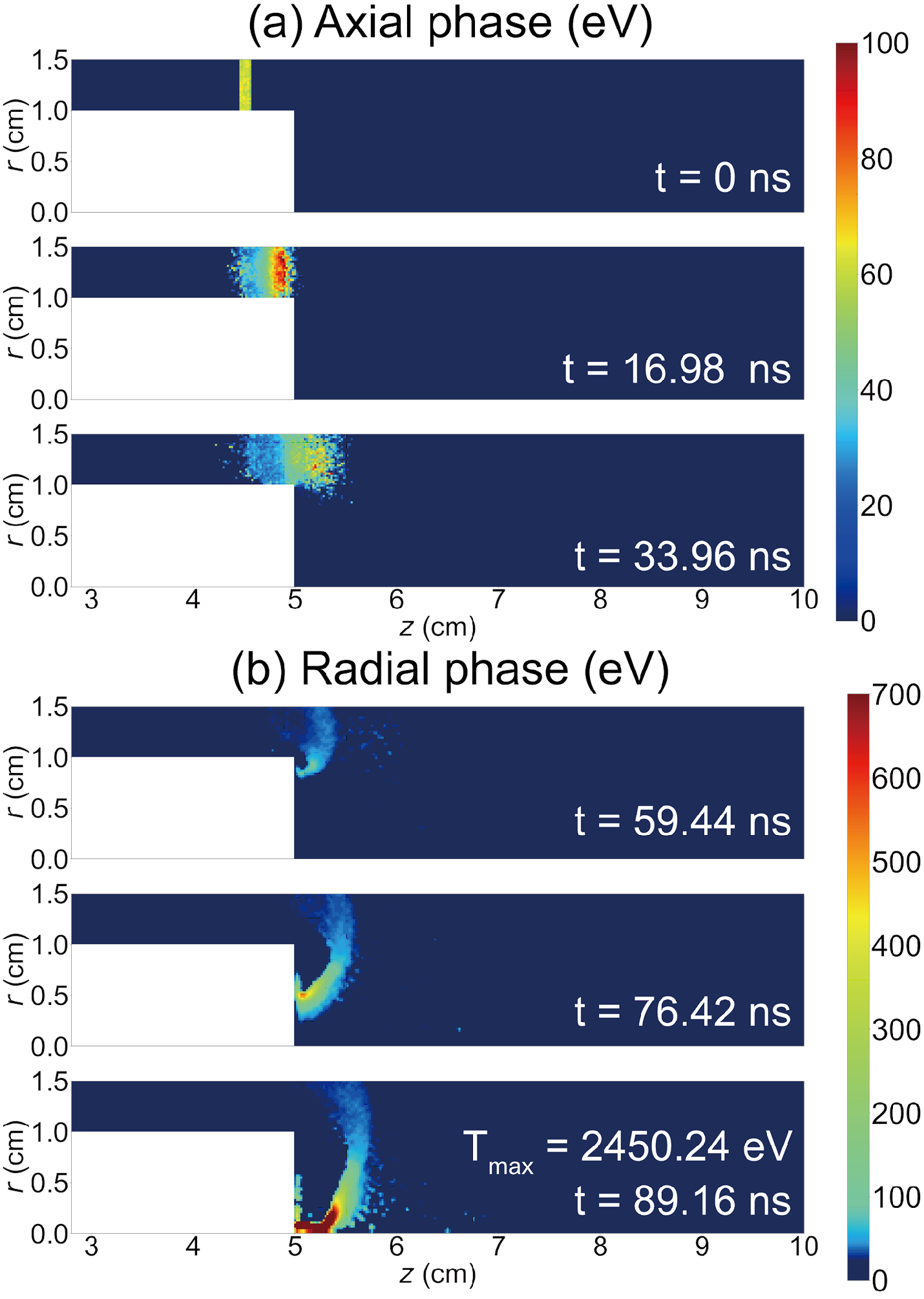}
    \caption{Ion temperature evolution in the hybrid simulation. Localized heating develops at the sheath front during the axial phase, and a compact hot region forms near the axis in the radial phase, where the maximum ion temperature reaches $T_{\max}\approx2.45\ \mathrm{keV}$. A second order median filter is applied to the ion temperature field before plotting in order to suppress cell scale particle noise while preserving the large scale structure.}
    \label{fig: ion temperature}
\end{figure}

\begin{figure}[htbp]
    \centering
    \includegraphics[width=0.5\linewidth]{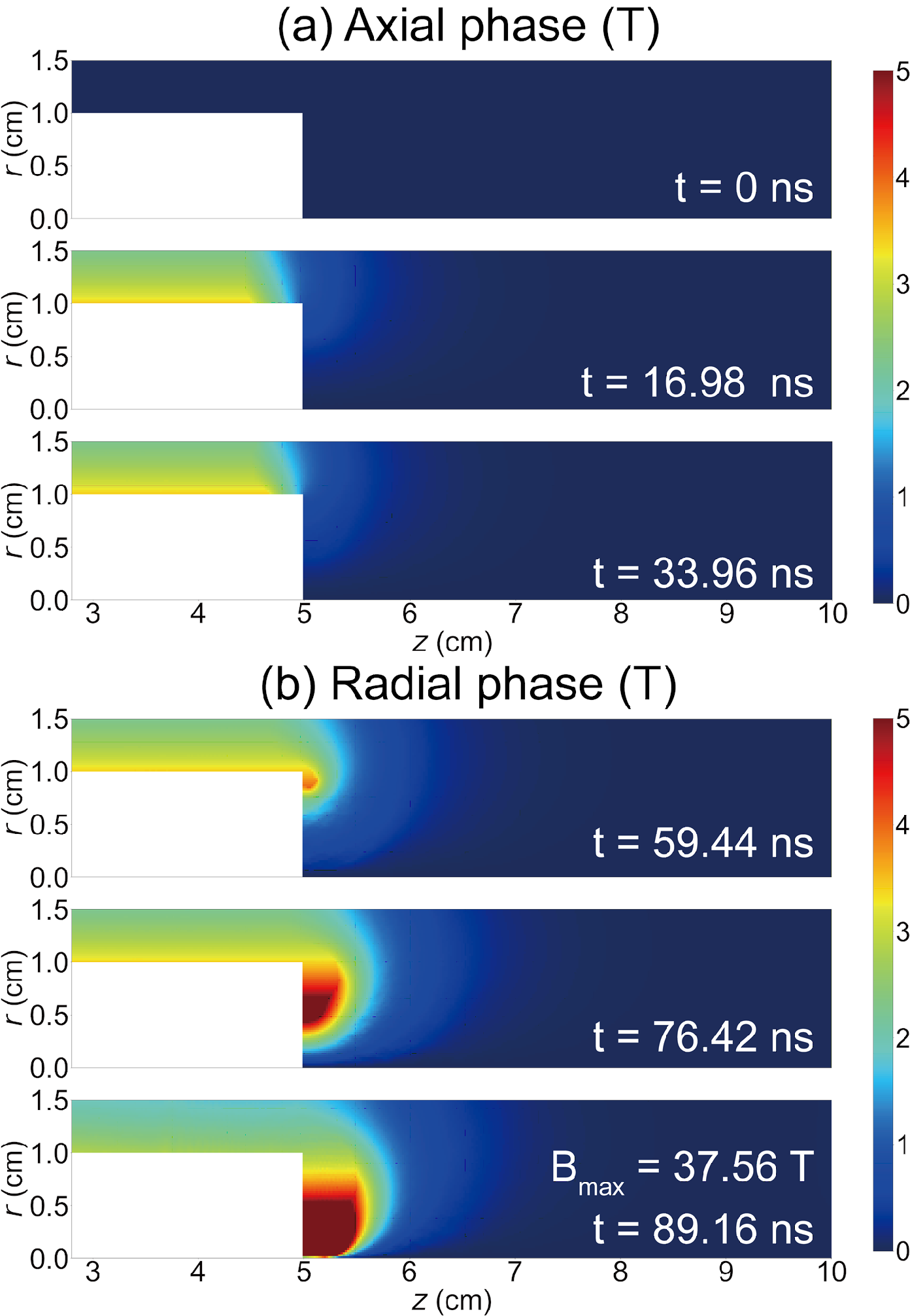}
    \caption{Evolution of the azimuthal magnetic field during the (a) axial and (b) radial phases. The self generated magnetic field builds up around the anode surface and tip, reaching a peak value $B_{\max}\approx37.6\ \mathrm{T}$ near the end of radial compression.}
    \label{fig: Ba}
\end{figure}

\begin{figure}[htbp]
    \centering
    \includegraphics[width=0.5\linewidth]{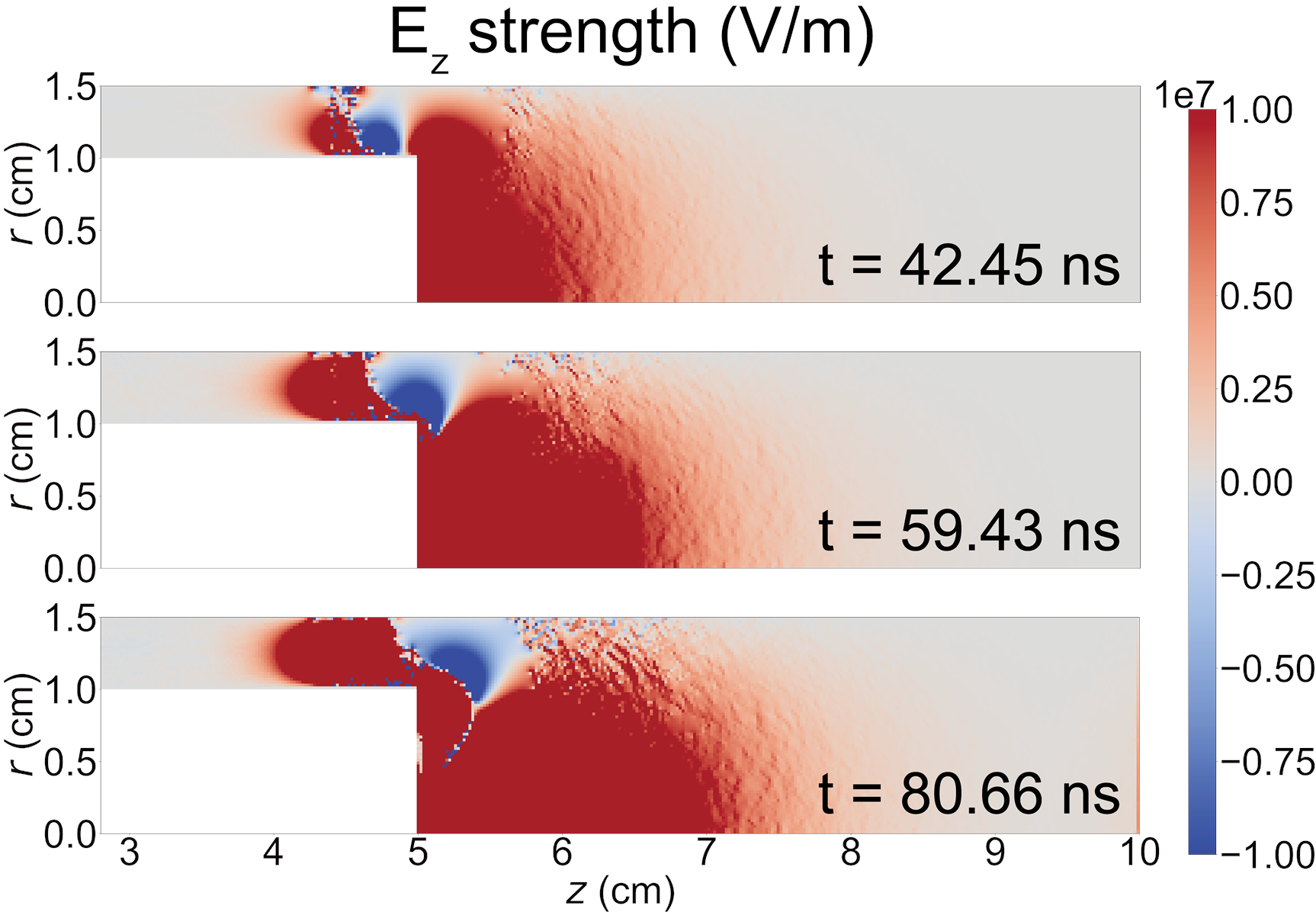}
    \caption{Axial electric field $E_z$ at representative times before, during, and after pinch. Strong bipolar fields form at the sheath front during rundown and compression, followed by an extended mainly positive field region in the post-pinch stage that drives beam like ion acceleration.}
    \label{fig: Ez}
\end{figure}

\begin{figure}[htbp]
    \centering
    \includegraphics[width=0.5\linewidth]{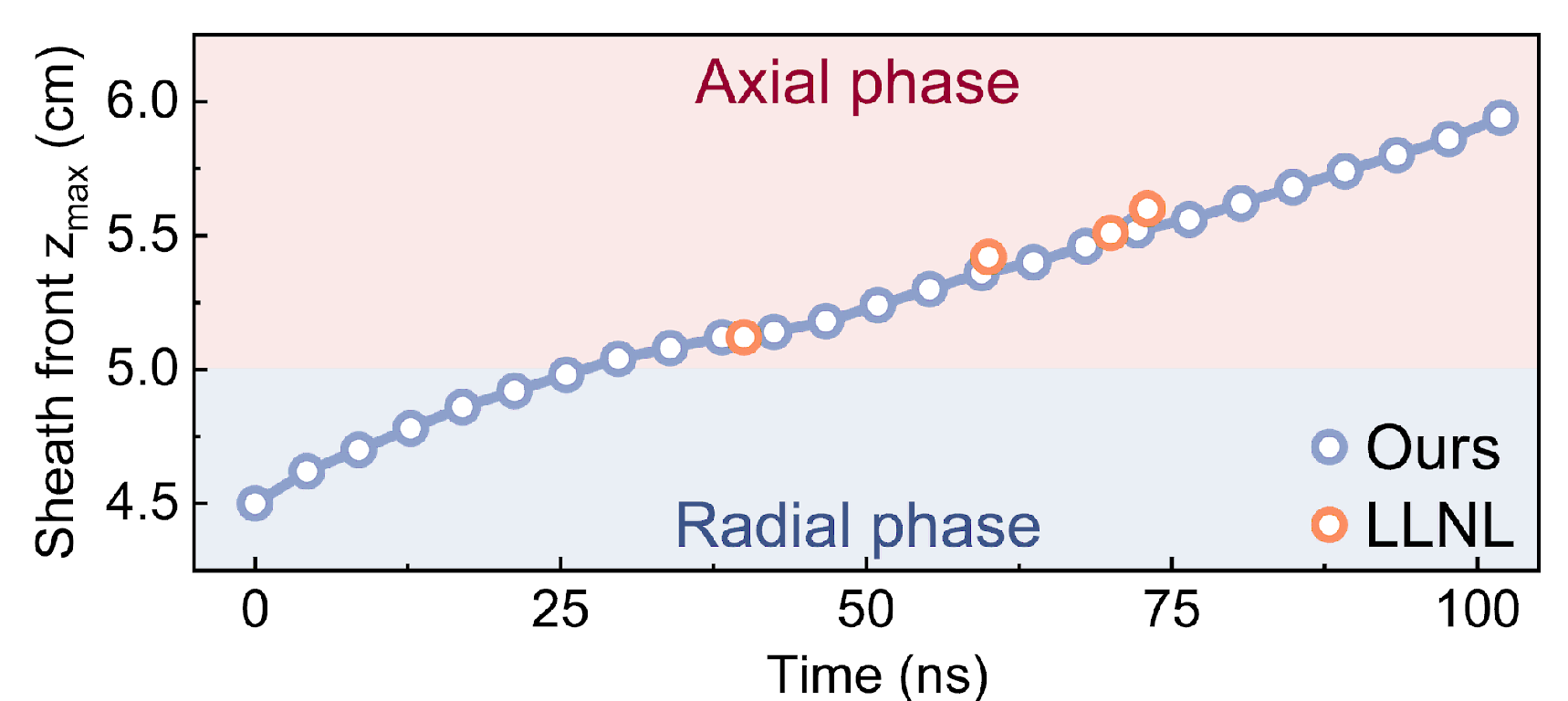}
    \caption{Sheath front position $z_{\max}$ as a function of time. Blue circles show the hybrid simulation result and orange markers indicate sheath positions inferred from LLNL fully kinetic simulations and measurements. Shaded bands distinguish axial and radial phases, demonstrating good agreement in sheath kinematics over the measured range.}
    \label{fig: sheath front}
\end{figure}

In this subsection we present the main discharge dynamics obtained from the default hybrid configuration described in Sec.\ref{subsec: simulation setup}. The analysis focuses on the evolution of ion density, ion temperature, magnetic field, and axial electric field, as shown in Fig.\ref{fig: sgm 001}-\ref{fig: Ez}. These results are used to characterize the transition from axial rundown to radial compression and pinch, and to provide the plasma parameters that enter the sheath front comparison in Fig.\ref{fig: sheath front} and Tab.\ref{tab: sheath compare} as well as the neutron yield evaluation.

% \begin{figure}[htbp]
%     \centering
%     \includegraphics[width=0.95\linewidth]{fig_ionDensity_baseline.png}
%     \caption{Baseline case (no electron‑pressure term, no Hall term, no predictor-corrector; “EHP000”): spatiotemporal evolution of ion number density $n_i(r,z,t)$. (a) axial‑phase snapshots during sheath propagation; (b) radial‑phase snapshots approaching pinch. Times (ns) are annotated on each panel.
% }
%     \label{fig: ion density}
% \end{figure}

Fig.\ref{fig: sgm 001} shows the time evolution of the ion number density during the discharge. At \(t = 0~\mathrm{ns}\), a thin (\(\sim 1~\mathrm{mm}\)) sheath with density \(n_s \approx 3.3\times10^{23}~\mathrm{m^{-3}}\) is initialized near \(z \approx 4.5~\mathrm{cm}\) on the anode surface, while the rest of the domain is filled with uniform background plasma at the prescribed density. At \(t = 16.98~\mathrm{ns}\) and \(t = 33.96~\mathrm{ns}\), the system is in the axial phase: the current-carrying sheath propagates along the anode toward increasing \(z\) under the action of the self-generated azimuthal magnetic field, maintaining a narrow high-density front that sweeps up and ionizes the background plasma.

Once the sheath reaches the anode tip, the discharge enters the radial phase. At \(t = 59.44~\mathrm{ns}\) the sheath begins to bend around the anode end and the dense front acquires an arched shape that is concave toward the axis. By \(t = 76.42~\mathrm{ns}\) the front has wrapped further around the tip and the high-density region is clearly converging toward smaller radii. At \(t = 89.16~\mathrm{ns}\) the sheath has almost closed around the tip and approaches \(r \approx 0\), signaling the onset of radial compression and the formation of a dense pinch column. The sequence in Fig.\ref{fig: sgm 001} therefore captures the characteristic transition from axial propagation to radial implosion that is expected for dense plasma focus discharges.

Fig.\ref{fig: ion temperature} shows the evolution of the ion temperature during the same discharge as in Fig.\ref{fig: sgm 001}. At \(t = 0~\mathrm{ns}\) the plasma outside the initialized sheath is close to the cold background state, and only a narrow band on the anode surface reaches tens of electron volts. As the axial phase develops \((t = 16.98~\mathrm{ns}\) and \(t = 33.96~\mathrm{ns})\), a localized temperature peak forms at the leading edge of the sheath. This hot region is confined near the anode surface and moves with the sheath front, indicating that electromagnetic energy is preferentially deposited in the vicinity of the moving current channel. For clarity, a second order median filter is applied to the temperature field before visualization, which smooths cell to cell fluctuations caused by finite macroparticle statistics without changing the overall temperature profile.

When the discharge enters the radial phase \((t = 59.44~\mathrm{ns}\) and \(t = 76.42~\mathrm{ns})\), the hot region bends around the anode tip together with the dense sheath and begins to contract toward the axis. The temperature distribution becomes more asymmetric, with a well defined hot spot on the inner side of the curved sheath. At \(t = 89.16~\mathrm{ns}\) the radial compression is nearly complete and a compact high temperature core forms close to \(r = 0\). The peak ion temperature reaches \(T_{\max} \approx 2.45~\mathrm{keV}\), while the surrounding plasma remains much cooler. This sequence demonstrates that the hybrid model captures strong, localized ion heating associated with sheath driven compression and pinch formation, which is essential for subsequent neutron production and energetic ion generation.

Fig.\ref{fig: Ba} shows the evolution of the azimuthal magnetic field around the anode during the discharge. At \(t = 0~\mathrm{ns}\) the field is essentially zero everywhere, since the external circuit current has not yet built up. As the current sheath forms and begins to propagate axially along the anode surface \((t = 16.98~\mathrm{ns}\) and \(t = 33.96~\mathrm{ns})\), a smooth band of \(B_\theta\) develops that is strongest near the anode and decays toward the vacuum region. The field pattern remains approximately straight along the anode, consistent with the predominantly axial motion of the current channel during this stage.

When the sheath reaches the anode tip and the discharge enters the radial phase \((t = 59.44~\mathrm{ns}\) and \(t = 76.42~\mathrm{ns})\), the region of strong \(B_\theta\) bends around the tip and begins to wrap toward the axis. The magnetic field maximum moves toward smaller radii and becomes more localized around the curved current channel, indicating an increasing magnetic pressure that drives radial compression. At \(t = 89.16~\mathrm{ns}\) the field reaches its peak magnitude, \(B_{\max} \approx 37.6~\mathrm{T}\), concentrated near the anode tip and close to the axis. This strong, localized azimuthal field provides the main confinement for the pinch column and is consistent with the high density and temperature achieved at the same time in Fig.\ref{fig: sgm 001} and Fig.\ref{fig: ion temperature}.

Fig.\ref{fig: Ez} shows the evolution of the axial electric field \(E_z\) at three representative times during the late rundown and radial compression phases. At \(t = 42.45~\mathrm{ns}\) the sheath has approached the anode tip and a localized bipolar structure in \(E_z\) forms at the sheath front, with a negative lobe on the upstream side and a positive lobe on the downstream side. Outside this narrow region the axial field remains relatively weak, indicating that the electric field is still mainly confined to the vicinity of the current sheath.

As the discharge evolves to \(t = 59.43~\mathrm{ns}\) the bipolar structure strengthens and extends along the curved sheath that wraps around the anode tip. The magnitude of \(E_z\) reaches the order of \(10^{7}~\mathrm{V/m}\), and a broader region of positive field develops downstream of the sheath, signalling enhanced conversion of magnetic energy into electrostatic potential. By \(t = 80.66~\mathrm{ns}\) the positive \(E_z\) region has further expanded in the axial direction, while a residual negative lobe remains near the collapsing sheath. This large scale, predominantly positive axial field is consistent with the onset of post-pinch acceleration and provides the main driving force for the formation of high energy ion beams observed in the simulations.

\begin{table}[htbp]
\centering
\begin{tabular}{cccc}
\hline
$z_{\max}$ /cm & \textbf{LLNL} /ns & \textbf{Ours} /ns & \textbf{Difference} /$\%$ \\ \hline
5.12          & 40        & 38.21     & -4.47            \\
5.42          & 60        & 65.39     & -8.99            \\
5.51          & 70        & 71.26     & 1.80            \\
5.60          & 73        & 79.42     & 8.79            \\ \hline
\end{tabular}
\caption{Axial locations of the outermost sheath front and the corresponding arrival times. The LLNL column lists arrival times reported in the reference case, while the values in the ``Ours'' column are obtained by linearly interpolating the hybrid simulation curve in Fig.\ref{fig: sheath front} at the same $z_{\max}$. The rightmost column gives the relative difference $(t_{\text{ours}}-t_{\text{LLNL}})/t_{\text{LLNL}}\times 100\%$.}
\label{tab: sheath compare}
\end{table}

To enable a consistent comparison between the present hybrid simulations and the fully kinetic LLNL results, we first specify how the sheath front position $z_{\max}(t)$ is defined. Let $n_i(r,z,t)$ denote the ion number density and $n_0$ the initial background density in the pre-filled region. During the run-down phase the outer current sheath propagates approximately along $z$ at a radius close to the anode surface, so we examine $n_i$ as a function of $z$ at that radius. At each time $t$ the sheath front $z_{\max}(t)$ is defined as the axial location of the first cell where the local ion density exceeds a fixed threshold,
\begin{equation}
    n_i(r,z_{\max}(t),t) \ge 1.5\,n_0.
\end{equation}

This density-based criterion provides an objective and reproducible way of extracting the sheath-front trajectory from the simulation data. In the LLNL study, by contrast, sheath-front positions were inferred by visually identifying the leading edge of the ion-density contours in the published plots and digitizing the corresponding coordinates. In the present comparison we therefore apply the $1.5\,n_0$ criterion to our hybrid results, while the LLNL data points follow their original, image-based definition.

Fig.\ref{fig: sheath front} summarizes the motion of the outermost current sheath by plotting the axial position of the sheath front, $z_{\max}$, as a function of time. The blue markers show the present hybrid simulation, where $z_{\max}$ is extracted from the ion-density maps using the fixed density threshold described above. The orange markers represent sheath-front positions inferred from the fully kinetic LLNL results. The shaded bands indicate the axial and radial phases of the discharge. During the axial phase the sheath advances almost linearly in time while remaining closely attached to the anode surface. After the front reaches the anode tip, the trajectory transitions into the radial phase, where the current channel bends around the tip, the effective propagation speed decreases, and the front continues to move downstream as it converges toward the axis.

To quantify the agreement with the LLNL benchmark, Tab.\ref{tab: sheath compare} compares the arrival time of the outermost sheath at four axial locations. For each \(z_{\max}\), the hybrid arrival time is obtained by linearly interpolating the simulated \(z_{\max}(t)\) curve in Fig.\ref{fig: sheath front}; the LLNL values are obtained in the same way from the digitized sheath-front trajectory reported in Ref.\cite{schmidt2012fully}. At \(z_{\max}=5.12~\mathrm{cm}\) the hybrid sheath arrives about \(1.79~\mathrm{ns}\) earlier than the LLNL result, corresponding to a relative difference of \(-4.47\%\). The largest deviation occurs at \(z_{\max}=5.42~\mathrm{cm}\), where the hybrid arrival time is later by \(5.39~\mathrm{ns}\) (\(8.99\%\)). At \(z_{\max}=5.51~\mathrm{cm}\) the difference is reduced to \(1.26~\mathrm{ns}\) (\(1.80\%\)), and at \(z_{\max}=5.60~\mathrm{cm}\) the sheath lags the LLNL value by \(6.42~\mathrm{ns}\) (\(8.79\%\)). Overall, the sheath-front kinematics predicted by the hybrid model agree with the fully kinetic reference to within about \(10\%\) over the range where comparison data are available, which is sufficient for the subsequent analysis of pinch dynamics and neutron production.

Subsequently, the neutron yield of the simulated non-hollow DPF configuration was calculated using the hybrid model. The results show that the neutron yield is predominantly released within a narrow time window during the transition between the pinch and post-pinch phases, corresponding to the period when both plasma density and temperature reach their maximum values. 

\begin{figure}[htbp]
    \centering
    \includegraphics[width=0.4\linewidth]{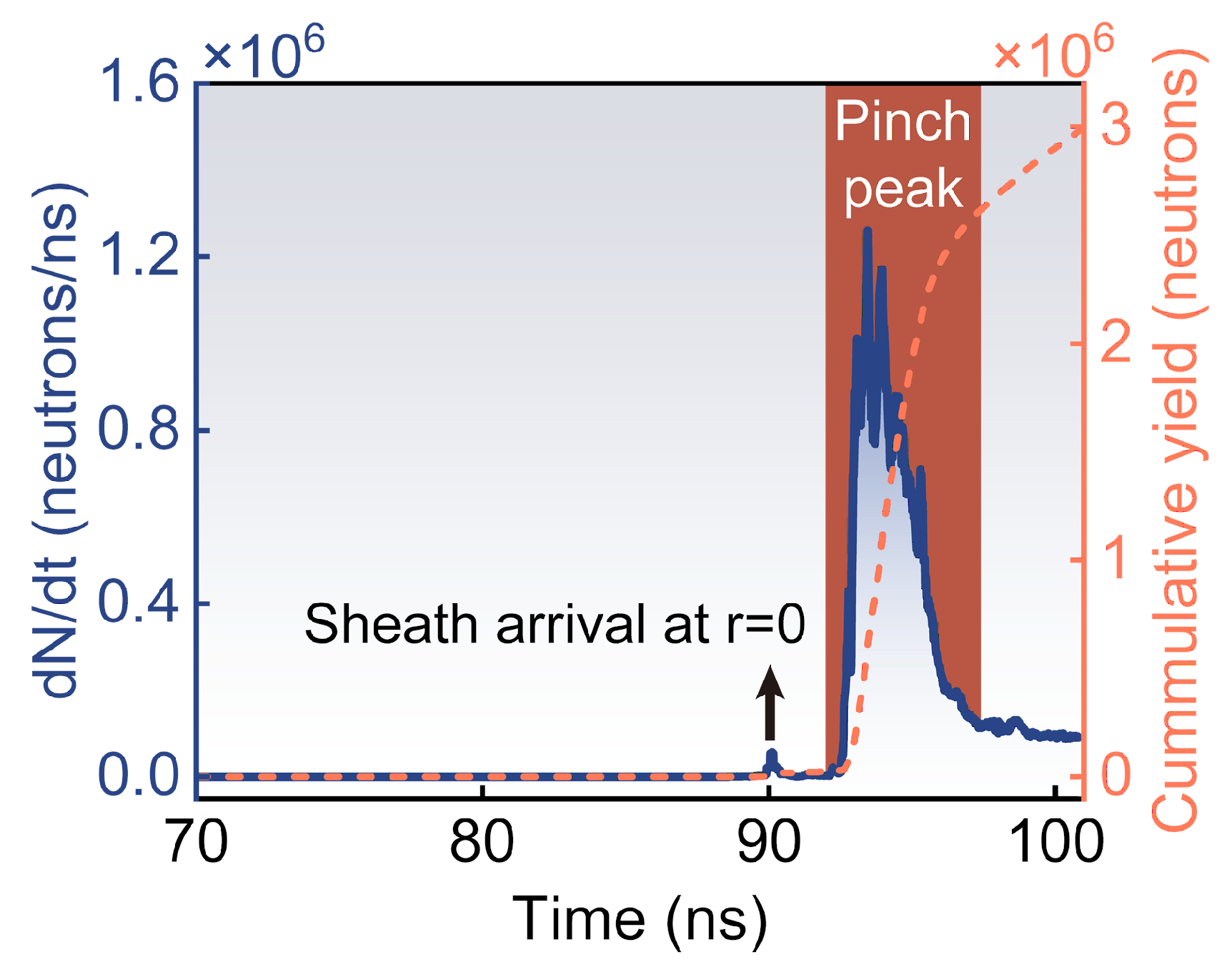}
    \caption{Time evolution of the instantaneous neutron production rate $\mathrm{d}N/\mathrm{d}t$ (solid blue, left axis) and cumulative yield $N(t)$ (dashed red, right axis). The arrow marks the time when the plasma sheath first reaches the axis at $r=0$, and the shaded region indicates the main pinch burst during which almost all neutrons are produced.}
    \label{fig:neutron_yield}
\end{figure}

Fig.\ref{fig:neutron_yield} shows that the neutron production rate remains negligible during the rundown phase and starts to rise only when the current sheath reaches the axis at $r=0$. Shortly after sheath arrival a very sharp burst develops: $\mathrm{d}N/\mathrm{d}t$ increases by many orders of magnitude, reaches its peak within a time of order $1~\mathrm{ns}$, and then decays rapidly, while the cumulative yield $N(t)$ approaches a plateau. More than $90\%$ of the total neutrons are generated inside the shaded pinch window, which confirms that the neutron emission is tightly localized around the pinch.

By integrating the neutron production rate over time, the total neutron yield obtained from the hybrid model simulation is $0.296\times10^{7}$, which is consistent with or even superior to typical results reported in previous numerical studies. For comparison, the work of A.~Schmidt \emph{et al.}\cite{schmidt2012fully} reported a neutron yield of approximately $0.86\times10^{7}$ for a hollow DPF configuration using a fully kinetic model, while their hybrid model yielded only $3.6\times10^{4}$. In contrast, the present hybrid model combined with a non-hollow geometry achieves a yield of the same order of magnitude as their fully kinetic simulation. This finding demonstrates that the hybrid approach employed in this work can achieve neutron yield prediction accuracy comparable to that of the fully kinetic model by A.~Schmidt \emph{et al.}\cite{schmidt2012fully}, while significantly outperforming their hybrid model results by several orders of magnitude.

The hybrid simulations reported here use the same outer electrode dimensions as the compact LLNL DPF, but with a solid anode rather than the hollow anode employed in the reference experiments and fully kinetic studies. In the LLNL configuration the hollow is designed to seed an $m=0$ instability and to enhance beam-target fusion in the post-pinch phase. Dedicated experimental campaigns have shown that, for a fixed bank energy and current, hollow anodes systematically produce higher neutron yields than solid ones. Optimizing the hollow diameter on a kilojoule-class DPF increased the average neutron yield by more than a factor of six compared with a non-hollow anode,\cite{shaw2018maximizing} and measurements on similar devices report maximum yields of order $10^{8}$ neutrons per shot for hemispherical hollow anodes.\cite{novotny2023effect} It is therefore reasonable to expect that a non-hollow anode operated at the same current level will produce a somewhat lower neutron yield than a hollow configuration, even if the global current waveform is similar. Moreover, detailed experimental data for a non-hollow, hundred-kiloampere DPF with exactly the same geometry as considered here are not publicly available, so the comparison with LLNL results should be interpreted at the level of order-of-magnitude agreement rather than as a strict one-to-one validation.

\begin{figure}[htbp]
    \centering
    \includegraphics[width=0.5\linewidth]{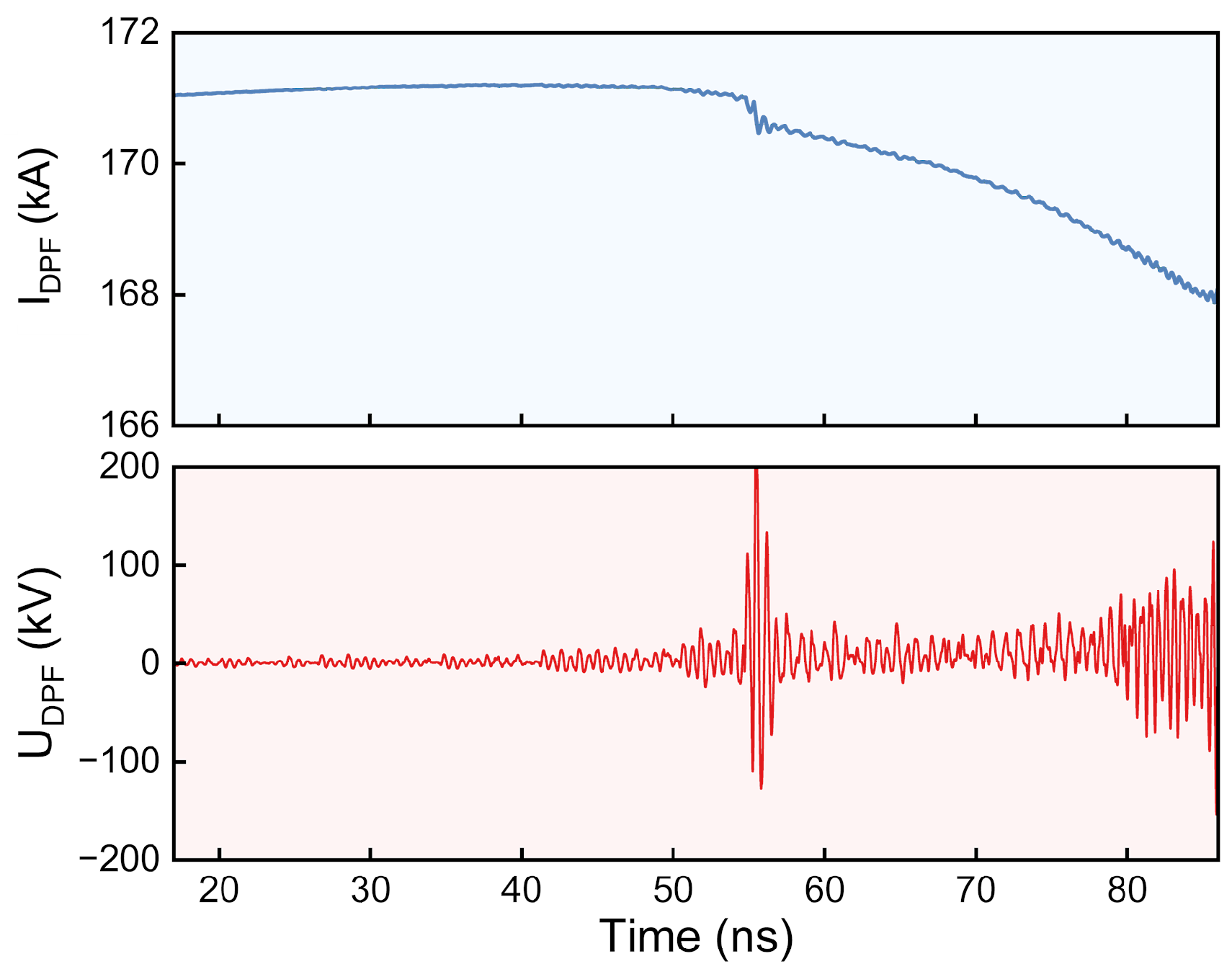}
    \caption{Time evolution of the discharge current $I_{\rm DPF}$ and system voltage $U_{\rm DPF}$ obtained from the hybrid simulation. The current between electrodes remains close to $180~\mathrm{kA}$ during the axial phase, then experiences a small drop at the onset of radial compression and gradually decreases in the post-pinch stage. The voltage across the plasma remains relatively small in the early phase, but develops large oscillations of order $10^{5}~\mathrm{V}$ around the time of pinch formation and during the subsequent relaxation.}
    \label{fig: iu}
\end{figure}

Fig.\ref{fig: iu} shows the time histories of the discharge current $I_{\rm DPF}$ and plasma voltage $U_{\rm DPF}$ for the same shot as in Sec.\ref{subsec: circuit and B}. The current in the external circuit rises rapidly to about $180~\mathrm{kA}$ and then stays nearly constant during the axial phase, indicating that most of the applied voltage is dropped across the static circuit elements. As the current sheath reaches the anode tip and the discharge enters the radial phase, a small but clear reduction in $I_{\rm DPF}$ appears and the current slowly decreases as magnetic energy is converted into plasma compression and heating. In contrast, the voltage across the plasma remains low and weakly oscillatory at early times, but develops a burst of strong oscillations with peak amplitude on the order of $10^{5}~\mathrm{V}$ around the time of pinch formation, followed by damped oscillations in the post-pinch stage. The correlation between the voltage spikes and the pinch phase is consistent with the rapid change of magnetic flux linked to the external circuit and provides an independent signature of the implosion and subsequent expansion.

Qualitatively, the simulated current trace exhibits the familiar features of dense plasma focus discharges: a fast rise from zero to a plateau near 180 kA during the axial phase, a modest reduction as the radial implosion converts magnetic energy into compression and heating, and a gradual decay in the post-pinch stage. These characteristics are consistent with semi-empirical models and experimental observations of compact DPF devices, where the rundown time and peak current are largely controlled by the external inductance and the fill pressure. Although we do not perform a systematic scan in $L_0$ or in the gas pressure here, the chosen configuration lies in the parameter regime of existing kilojoule-class DPF experiments, so the simulated current waveform can be regarded as representative of this class of devices.

These results clearly indicate that a properly constructed hybrid model can provide both high accuracy and high computational efficiency. This establishes a promising technical pathway for future high-fidelity and cost-effective numerical simulations of DPF devices.

\subsection{Resolution and numerical parameter sensitivity}
\label{subsec: para sensitivity}

Having established the baseline hybrid PIC fluid configuration and validated its ability to reproduce the main phases of the DPF evolution, we next examine the sensitivity of the results to numerical resolution and key model parameters. In particular, we focus on the sheath front trajectory and neutron yield, since these are central to the comparison with fully kinetic simulations and experiments, and they are also the most likely to be affected by choices such as grid spacing, time step, conductivity thresholds, divergence cleaning strength, and the electron temperature closure.

\begin{figure}[htbp]
    \centering
    \includegraphics[width=0.5\linewidth]{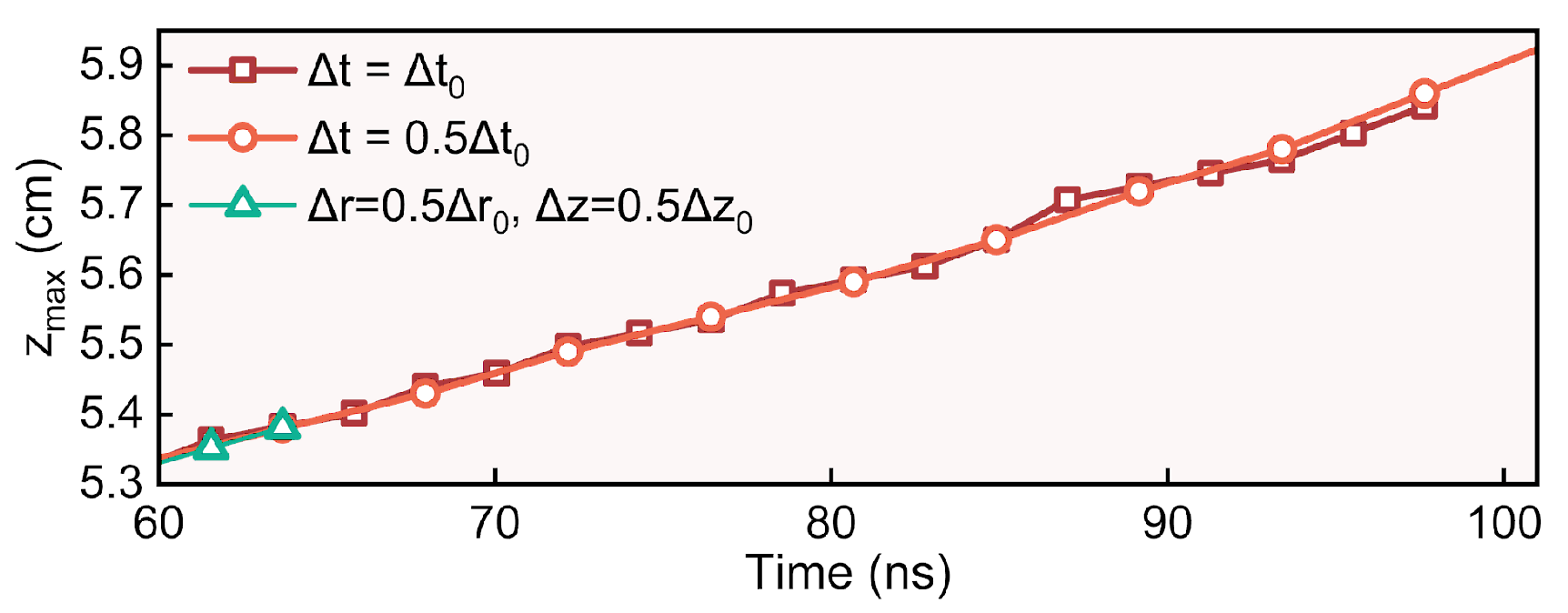}
    \caption{Sheath front position $z_{\max}(t)$ for the baseline run with $\Delta t=\Delta t_{0}$, a refined time step run with $\Delta t=0.5\Delta t_{0}$, and a refined spatial resolution run with $\Delta r=0.5\Delta r_{0}$ and $\Delta z=0.5\Delta z_{0}$, showing that the macroscopic sheath motion is insensitive to further refinement in space and time.}
    \label{fig: drdzdt}
\end{figure}

Fig.\ref{fig: drdzdt} compares the sheath front trajectory \(z_{\max}(t)\) for the baseline run, a run with halved time step \(\Delta t = 0.5\Delta t_0\), and a run with both spatial steps halved \(\Delta r = 0.5\Delta r_0\), \(\Delta z = 0.5\Delta z_0\) (which also satisfies the CFL condition with \(\Delta t = 0.5\Delta t_0\)). Up to \(t \simeq 100~\mathrm{ns}\), where the PML remains well behaved, the three trajectories are almost indistinguishable: sheath arrival times at several reference axial positions differ by less than \(0.5\%\). For the refined time step run the total neutron yield is \(Y = 0.287\times10^{7}\), compared with \(0.296\times10^{7}\) for the baseline case, a difference of about \(3\%\). These results demonstrate that both the spatial and temporal resolutions used in the baseline configuration are sufficient for resolving the macroscopic sheath motion and neutron production, and that the refined runs mainly confirm the robustness of the solution rather than revealing unresolved dynamics.

\begin{figure}[htbp]
    \centering
    \includegraphics[width=0.5\linewidth]{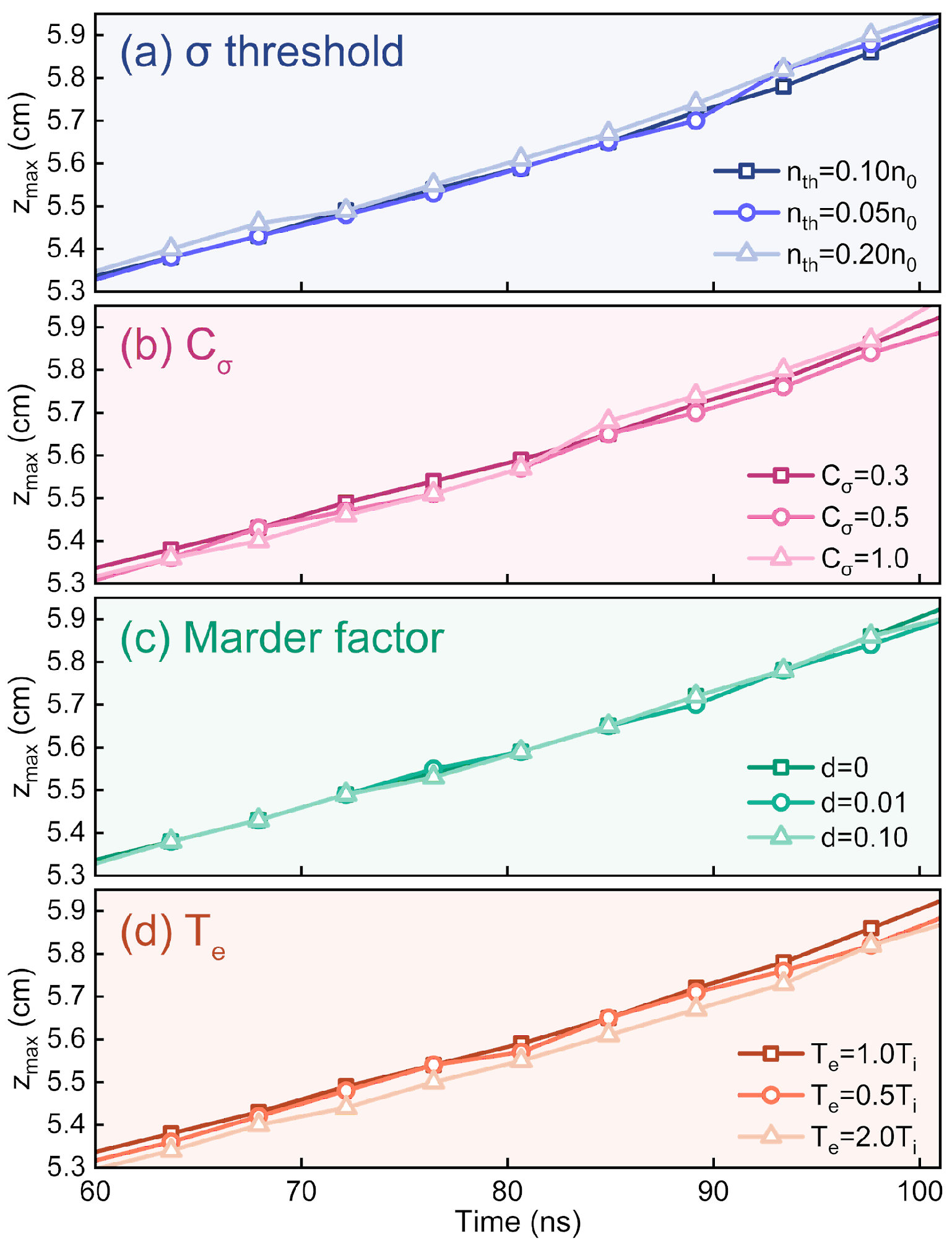}
    \caption{Sheath-front position $z_{\max}(t)$ for parameter scans: (a) conductivity threshold $n_{th}$, (b) Ohmic CFL factor $C_{\sigma}$, (c) Marder factor $d$, and (d) electron temperature closure $T_e = \alpha T_i$. The trajectories are nearly identical, indicating weak sensitivity to these parameters.}
    \label{fig: para sensitivity}
\end{figure}

To assess whether the inferred sheath dynamics is robust with respect to the numerical and modeling choices in the hybrid formulation, we perform a set of parameter sensitivity tests summarized in Fig.\ref{fig: para sensitivity}. Each panel varies one parameter that either enters the conductivity model, controls a stability limiter, or specifies a closure for the electron fluid. The aim is to verify that the sheath motion is not an artifact of these particular choices and to quantify the residual uncertainty they introduce.

Fig.\ref{fig: para sensitivity}(a) shows the evolution of the sheath front position \(z_{\max}(t)\) for different conductivity thresholds \(n_{\mathrm{th}}\) in \(\sigma(n_e)\). The threshold controls how rapidly the medium transitions from vacuum to a highly conducting plasma near the sheath front and could in principle affect the axial acceleration of the sheath. In the interval \(60~\mathrm{ns} \le t \le 100~\mathrm{ns}\) the three trajectories for \(n_{\mathrm{th}} = 0.05 n_0\), \(0.10 n_0\), and \(0.20 n_0\) almost coincide: the spread in \(z_{\max}\) remains below \(0.04~\mathrm{cm}\), corresponding to a relative difference of less than \(1\%\). This shows that moderate changes in the conductivity onset do not significantly modify the sheath propagation and that the reported sheath kinematics is not controlled by the particular choice of \(n_{\mathrm{th}}\).

Fig.\ref{fig: para sensitivity}(b) shows \(z_{\max}(t)\) for different Ohmic CFL factors \(C_{\sigma} = 0.3,\,0.5,\) and \(1.0\), which scale the maximum effective conductivity imposed by the stability limiter. In principle, a stronger limiter could artificially reduce the current and thus slow the sheath. Over \(60~\mathrm{ns} \le t \le 100~\mathrm{ns}\) the three curves remain nearly identical: the maximum spread in \(z_{\max}\) is below \(0.09~\mathrm{cm}\), about \(1.5\%\). This indicates that, in the range considered here, the CFL limiter mainly acts as a numerical safeguard that is only weakly active and does not control the global sheath motion.

Fig.\ref{fig: para sensitivity}(c) shows \(z_{\max}(t)\) for different Marder factors \(d = 0\), \(0.01\), and \(0.1\) used in the divergence cleaning of the electric field. Since this correction directly modifies \(\mathbf{E}\), it could in principle affect the sheath acceleration if chosen too large. In the interval \(60~\mathrm{ns} \le t \le 100~\mathrm{ns}\) the three trajectories are almost indistinguishable: the maximum spread in \(z_{\max}\) is below \(0.02~\mathrm{cm}\), corresponding to a relative difference of about \(0.3\%\). Combined with the observation that nonzero \(d\) slightly damps the peak fields (not shown), this confirms that the global sheath motion is essentially insensitive to the choice of \(d\) in this range and justifies using \(d = 0\) in the production runs.

Fig.\ref{fig: para sensitivity}(d) shows \(z_{\max}(t)\) for different electron temperature closures \(T_e = \alpha T_i\) with \(\alpha = 1\), \(0.5\), and \(2\). The electron temperature enters both the Spitzer conductivity and the electron pressure term, so variations in \(T_e\) provide a simple way to estimate the impact of the \(T_e = T_i\) closure on the sheath dynamics. In the interval \(60~\mathrm{ns} \le t \le 100~\mathrm{ns}\) the three trajectories remain very close: the maximum spread in \(z_{\max}\) is below \(0.06~\mathrm{cm}\), less than about \(1\%\). The \(T_e = T_i\) case is marginally ahead, while \(T_e = 2T_i\) lags slightly and \(T_e = 0.5T_i\) lies in between, consistent with the expected trend that higher \(T_e\) increases the conductivity and slightly reduces the effective accelerating field. These results indicate that the global sheath motion is only weakly sensitive to the precise value of \(T_e\) in the regime considered and that the \(T_e = T_i\) closure is adequate for the macroscopic observables analyzed in this work.

For completeness we have also recomputed the neutron yield for the three closures $T_{e}=\alpha T_{i}$ shown in Fig.\ref{fig: para sensitivity}(d). While the sheath trajectories differ by less than $1\%$ over $60\ \mathrm{ns}\le t\le100\ \mathrm{ns}$, the total neutron yield is more sensitive to $\alpha$, because fusion reactions are dominated by the high energy tail of the ion distribution, which is indirectly affected by the electron temperature through the conductivity and Ohmic heating. For $\alpha=0.5$ and $\alpha=2.0$ we obtain total yields of $2.86\times10^{5}$ and $5.30\times10^{5}$ neutrons, respectively, compared with $2.96\times10^{6}$ neutrons in the baseline case with $T_{e}=T_{i}$. These runs are therefore best interpreted as a parametric study of closure uncertainty rather than as alternative quantitative predictions of the neutron yield.

The values $\alpha=0.5$ and $\alpha=2.0$ are chosen to bracket plausible departures from ion-electron thermal equilibrium in DPF-like $Z$-pinches. Measurements and kinetic simulations of $Z$-pinches and dense plasma focus pinches indicate that the electron and ion temperatures can differ by factors of order unity, with both $T_{e}<T_{i}$ and $T_{e}>T_{i}$ observed depending on collisionality and heating mechanisms. Parameterized prescriptions for the electron-to-ion temperature ratio in the range of a few are also widely used in hybrid and GRMHD modelling when the fraction of dissipated energy that heats electrons is uncertain. Scanning over $\alpha\in[0.5,2]$ thus provides a simple way to estimate the impact of this uncertainty on macroscopic observables without introducing a separate electron energy equation.

Taken together, Fig.\ref{fig: drdzdt} and \ref{fig: para sensitivity} show that the baseline choices of spatial and temporal resolution and of the numerical and modeling parameters in the generalized Ohm law do not control the macroscopic discharge dynamics. The sheath front trajectories remain stable under substantial changes in grid spacing, time step, conductivity transition, CFL limiter, Marder factor, and electron temperature, with deviations at the level of a few percent and neutron yields differing by only a few percent between the baseline and refined runs. The dominant uncertainties in the present simulations therefore arise from the underlying physical approximations of the hybrid model rather than from the numerical parameter choices, and the baseline configuration can be regarded as numerically converged for the macroscopic observables of interest.

\subsection{Electron pressure and Hall contributions}

This subsection is not intended to provide a detailed physical study of electron pressure gradient and Hall effects in dense plasma focus discharges. Its primary purpose is to demonstrate that the numerical framework can stably incorporate these additional terms in the generalized Ohm law and to illustrate their qualitative influence on the sheath structure and pinch timing.

\begin{figure}[htbp]
    \centering
    \includegraphics[width=0.5\linewidth]{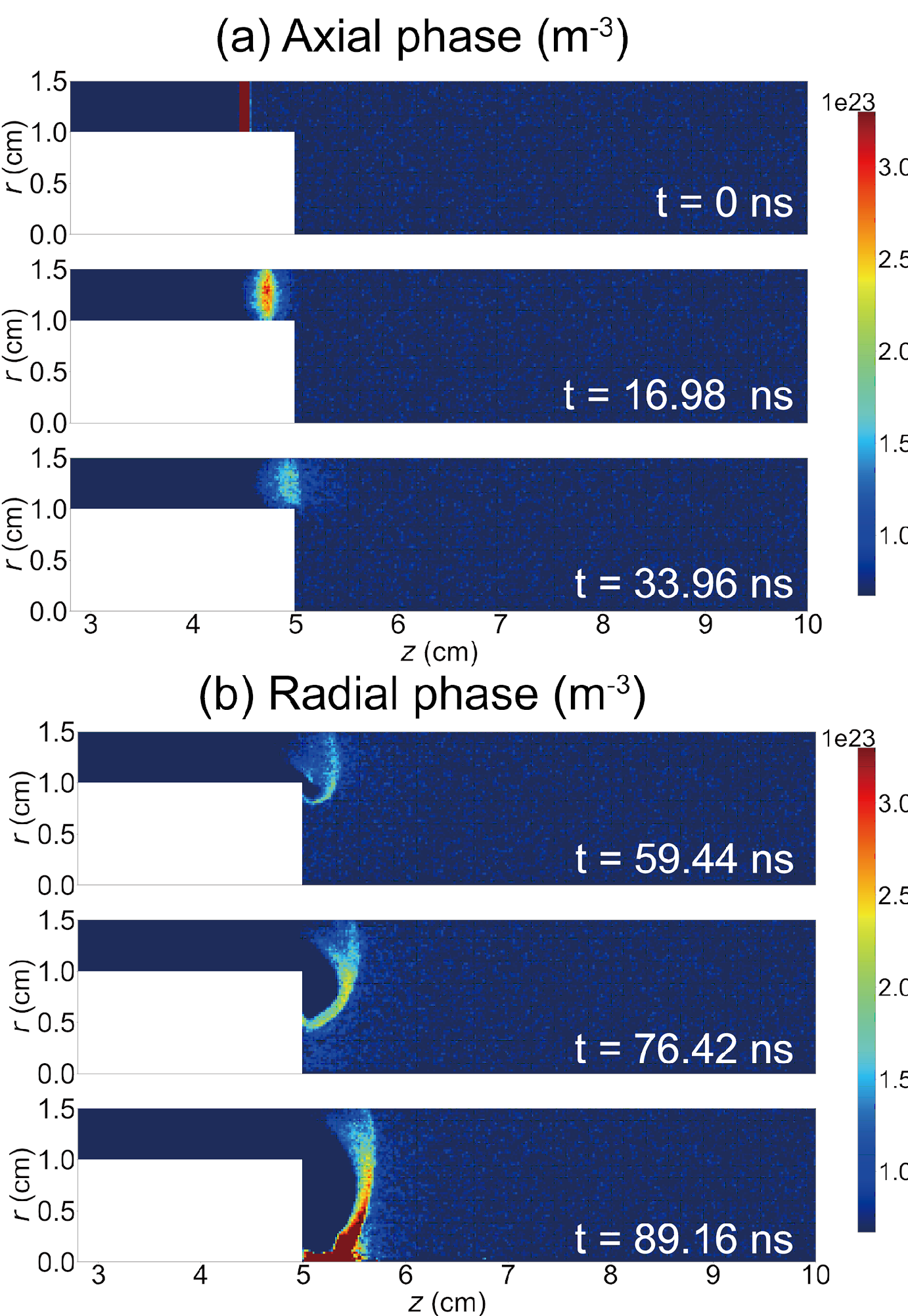}
    \caption{Time evolution of the ion number density when the electron-pressure gradient term is included in the generalized Ohm’s law. The sheath front propagates slightly faster, the pinch occurs earlier, and the high-density region becomes broader compared with the reference case without the electron-pressure gradient and Hall terms.}
    \label{fig: 101}
\end{figure}

\begin{figure}[htbp]
    \centering
    \includegraphics[width=0.5\linewidth]{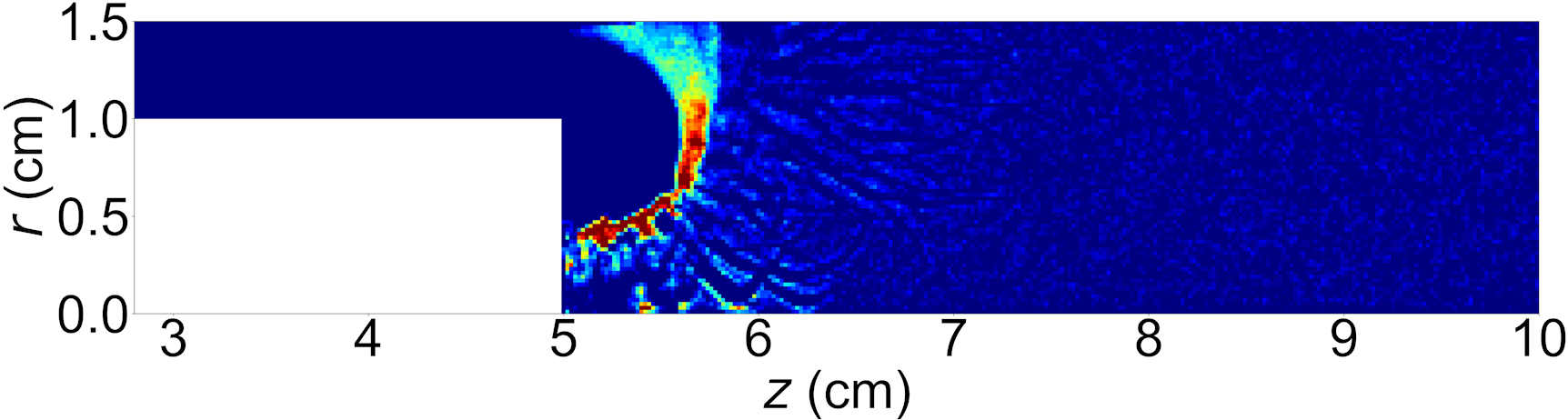}
    \caption{High-speed ion macroparticles and the ray-like structures produced when the electron pressure gradient is applied everywhere. In the final simulations, the pressure-gradient term is limited to cells with $n_i > 1.2\times10^{23}~\mathrm{m^{-3}}$ to avoid this effect while keeping the main plasma behavior.}
    \label{fig: high speed particle}
\end{figure}

\begin{figure}[htbp!]
    \centering
    \includegraphics[width=0.5\linewidth]{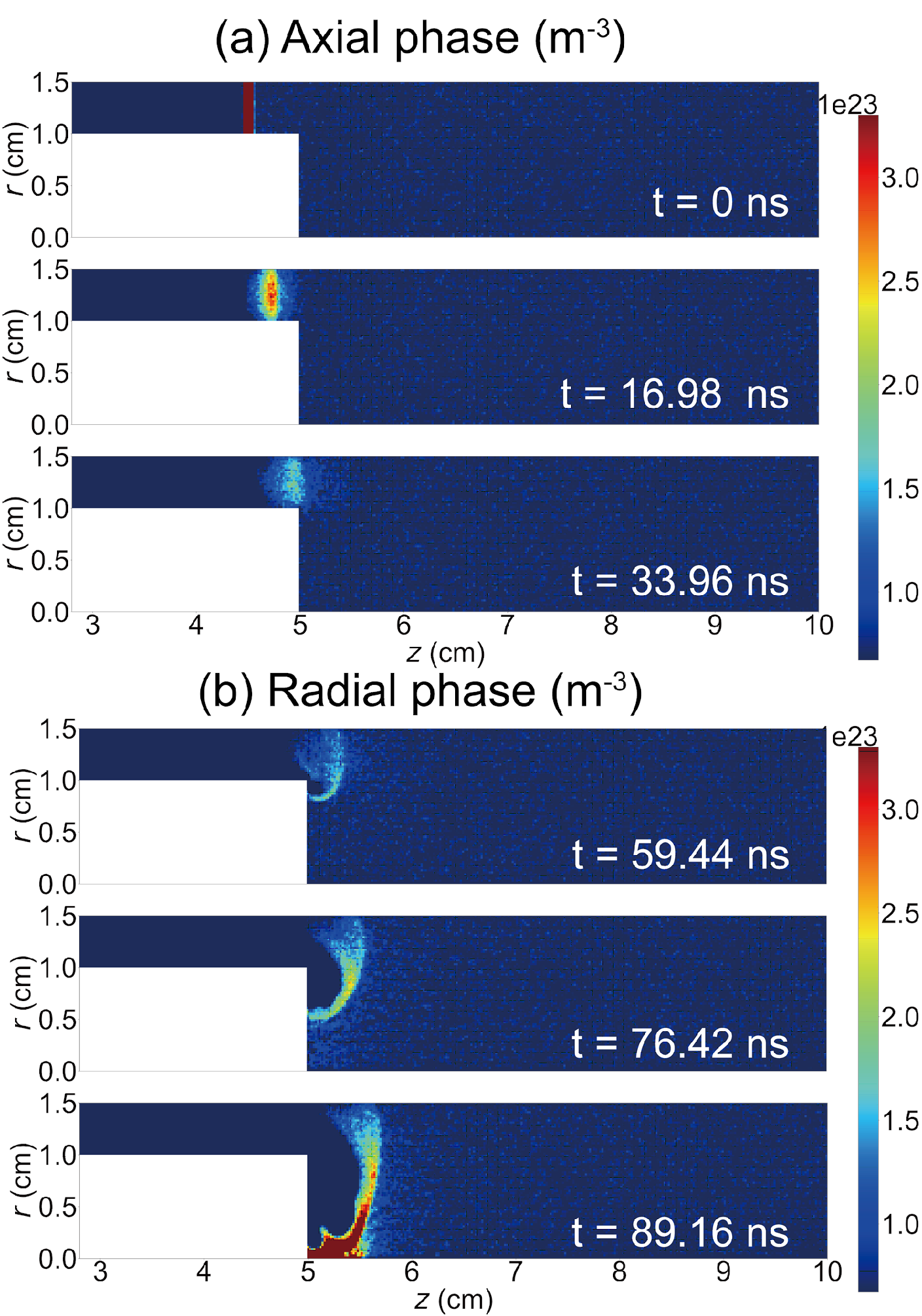}
    \caption{Time evolution of the ion number density when both the electron-pressure gradient and Hall terms are included in the generalized Ohm’s law. The Hall contribution further modifies the sheath dynamics, enhancing early-stage acceleration while amplifying small-scale structure and late-stage instability relative to the case without the Hall term.}
    \label{fig: 111}
\end{figure}

In this subsection we examine how the additional terms in the generalized Ohm law modify the sheath dynamics and pinch formation in the hybrid model. Starting from the reference case where only the resistive term and the $\mathbf{v}_e \times \mathbf{B}$ contribution are retained, we first activate the electron pressure gradient term and study its influence on the ion density evolution, as shown in Fig.\ref{fig: 101}. We then discuss a numerical difficulty that appears when the pressure gradient is applied in very low density regions and how this is mitigated by restricting the term to cells above a density threshold, illustrated in Fig.\ref{fig: high speed particle}. Finally, we add the Hall contribution and compare the resulting discharge behavior with the pressure only case in Fig.\ref{fig: 111}, in order to demonstrate that the code can handle the full generalized Ohm law and to assess the impact of these terms on sheath motion and pinch structure.

Fig.\ref{fig: 101} illustrates the evolution of the ion number density when the electron pressure gradient term in the generalized Ohm law is included while the Hall term is omitted. The initial sheath formation, axial rundown, and subsequent radial compression follow the same global sequence as in the reference case, but the sheath front advances slightly faster and the high density layer becomes broader in both the axial and radial phases. In particular, the dense front reaches the anode tip earlier and the peak density during radial convergence appears at an earlier time, indicating that the pressure gradient term provides an additional drive that enhances axial energy coupling and accelerates the overall implosion. These results confirm that the implementation of the \(\nabla p_{e}/(e n_{e})\) contribution is numerically robust and has a modest but systematic impact on sheath kinematics and pinch morphology.

Fig.\ref{fig: high speed particle} illustrates a numerical artefact that appears when the electron pressure gradient is evaluated throughout the entire domain. In this case, a set of macroparticles in the sheath region acquires anomalously large velocities and streams away from the dense front, producing a series of ray-like streaks that extend far into the low-density background. Because the average number of background macroparticles is only of order a few tens per cell, a single over-accelerated particle can strongly perturb the local fields and trigger a cascade in neighbouring cells, eventually driving the solution toward instability.

To suppress this nonphysical behaviour while retaining the beneficial influence of the pressure term on the sheath dynamics, we restrict the evaluation of the electron pressure gradient to cells with ion density above \(n_i > 1.0\times10^{23}~\mathrm{m^{-3}}\). With this localized treatment the spurious high-speed rays disappear, the simulation remains stable to late times, and the main features of sheath propagation and pinch formation are preserved.

In practice we set the density threshold to $n_{\mathrm{th}} = 1.0\times10^{23}\ \mathrm{m}^{-3}$. This corresponds to approximately $1.5\,n_{0}$, where $n_{0} = 6.7\times10^{22}\ \mathrm{m}^{-3}$ is the prefill background density listed in Table~I, and it matches the density contour used in Sec.IV~A to define the outer sheath front. With this choice, the electron pressure term is active only inside the compressed current sheath and pinch plasma, where the fluid description is expected to be most reliable, and is suppressed in the tenuous background and vacuum regions where the macroparticle statistics are poor and kinetic effects dominate. Varying $n_{\mathrm{th}}$ within a factor of two around this value mainly changes the thickness of the region in which the pressure gradient acts and the amount of spurious ray like structure that is removed, while the global sheath trajectory and the integrated neutron yield remain qualitatively unchanged. For the production runs we therefore adopt $n_{\mathrm{th}} = 1.0\times10^{23}\ \mathrm{m}^{-3}$ as a compromise between numerical stability and physical fidelity.

Fig.\ref{fig: 111} shows the corresponding ion number density evolution when both the electron pressure gradient and Hall terms are retained in the Ohm solver. The discharge again exhibits the expected transition from axial to radial motion, but the sheath front becomes sharper and more structured, especially during the late radial phase where small scale modulations appear along the dense layer. The peak density is slightly higher and the compression region is more concentrated than in the pressure-only case, which reflects the stronger local electromagnetic coupling introduced by the Hall contribution. At the same time, the predictor-corrector step suppresses the most violent numerical instabilities so that the simulation can be advanced through the pinch and early post-pinch stages. Overall, this case demonstrates that the code can self consistently handle the full generalized Ohm law, while highlighting that the Hall term tends to increase sensitivity to temperature and density gradients in the sheath region.

In summary, the extended Ohm law studies show that the electron pressure gradient and Hall terms mainly affect the detailed shape and timing of the current sheath rather than its overall trajectory. For the purposes of the present work they are therefore best regarded as qualitative demonstrations of code capability. All quantitative comparisons with fully kinetic benchmarks and with neutron yield estimates are based on the simpler configuration without these terms, which avoids over-interpreting electron-scale physics that is not yet constrained by a dedicated temperature model.

\subsection{Computational cost}

From a computational point of view, a fully kinetic electromagnetic PIC simulation of the same dense plasma focus discharge would be extremely expensive. In a fully kinetic description both ions and electrons are treated as particles and the grid spacing must resolve the Debye length and the electron skin depth. For typical pinch conditions with electron densities on the order of $n_e \sim 10^{24}$ to $10^{25}\,\mathrm{m^{-3}}$ and electron temperatures $T_e \sim 10^{2}\,\mathrm{eV}$, the Debye length is only a few micrometres, about two orders of magnitude smaller than the present cell size $\Delta r = \Delta z = 2\times 10^{-4}\,\mathrm{m}$. Resolving these microscopic scales in both $r$ and $z$ would increase the number of grid cells by roughly three to four orders of magnitude even in two dimensions, while the number of macroparticles would grow by at least one to two orders of magnitude due to the inclusion of electrons. Together with the stricter time step required to resolve the electron plasma period, this implies that a fully kinetic EM PIC simulation of the same configuration would be about $10^{5}$ to $10^{6}$ times more expensive than the present hybrid model in terms of particle updates and floating point operations.

In the present hybrid PIC-fluid model the grid consists of $78 \times 523$ cells and the plasma is represented by $5\times 10^{5}$ ion macroparticles in the main volume plus $2.6\times 10^{4}$ macroparticles in the pre-ionized layer near the anode. The code is parallelized with OpenMP. On an AMD EPYC 7542 32-core CPU using 16 OpenMP threads, the compute part of one explicit time step (excluding diagnostic output) requires approximately $0.084$ to $0.128\,\mathrm{s}$ of wall-clock time. For case in \ref{subsec: result}, the ion-density frame at $t = 89.16\,\mathrm{ns}$ corresponds to time step $N_t = 2.4\times 10^{5}$, so a complete discharge simulation up to this time requires about $(0.084$ to $0.128)\times 2.4\times 10^{5}\,\mathrm{s} \approx (2.0$ to $3.1)\times 10^{4}\,\mathrm{s}$ of wall-clock time, that is roughly $5.6$ to $8.5$ hours on 16 cores. This level of efficiency makes it feasible to simulate entire discharges and to carry out limited parameter scans, whereas a fully kinetic EM PIC treatment of the same dense plasma would be prohibitively expensive.

\subsection{Model limitations}

The present simulations employ a hybrid PIC-fluid description in which ions are treated kinetically while electrons are modelled as a quasineutral fluid. This reduction brings substantial savings in computational cost but introduces several limitations. First, the electron response is described by a scalar pressure and a simplified collisional conductivity; pressure anisotropy, non-Maxwellian tails, anomalous resistivity and turbulence-driven transport are not resolved. Second, quasineutrality is enforced and Debye sheaths and thin space-charge layers are not resolved, so near-wall sheath microphysics and microscopic double layers are outside the scope of the model. Third, the implementation is restricted to an axisymmetric two-dimensional geometry and therefore only the $m = 0$ mode is represented. Dense plasma focus pinches are known to exhibit $m = 1$ kink and higher-order azimuthal instabilities, which can shorten the pinch lifetime and redistribute the density. The present results should therefore be interpreted as providing macroscopic trends and integral quantities rather than a fully detailed description of three-dimensional pinch fragmentation.

Including the electron pressure gradient and Hall term in Ohm's law makes the solution more sensitive to the electron temperature field. In the current implementation the electron temperature is approximated as equal to the ion temperature, $T_e = T_i$, and no separate electron energy equation is solved. This closure is adequate for the baseline hybrid runs without $\nabla p_e$ and Hall contributions, but it becomes less accurate once these terms are retained, because $T_e$ directly enters the pressure gradient, the conductivity and hence the local electric field. As a consequence, simulations that include $\nabla p_e$ and the Hall term may exhibit additional discrepancies with experimental trends and fully kinetic benchmarks, especially in regions with strong temperature gradients and reduced collisionality. A natural next step is to introduce a separate evolution equation for $T_e$ with appropriate collisional coupling and heat-flux models, which is expected to improve the consistency of simulations that rely on the extended Ohm's law.Consequently, the simulations that include the $\nabla p_{e}$ and Hall terms under the present $T_{e}=T_{i}$ closure should be regarded mainly as qualitative explorations of how the extended Ohm law modifies sheath structure and pinch timing, rather than as quantitatively accurate predictions of electron-scale physics. A simple parameter scan with $T_e = \alpha T_i$, discussed in Sec.\ref{subsec: para sensitivity} indicates that this closure leaves the sheath trajectory essentially unchanged but introduces at least a factor of a few uncertainty in the absolute neutron yield.

\section{Conclusion}
\label{sec:conclusion}

In this work, we have developed an electromagnetic hybrid fluid-particle framework that couples kinetic ions with a fluid electron model and advances the full set of Maxwell equations, rather than using a Darwin or quasistatic approximation. This treatment allows the self consistent evolution of fields in both the dense plasma and the surrounding vacuum region while keeping the quasi neutral ion-electron closure and an efficient 2D axisymmetric geometry.

The simulations reproduce the typical four stage evolution of a dense plasma focus discharge, namely sheath formation, axial rundown, radial convergence, and post-pinch expansion. The time resolved ion density, ion temperature, magnetic field, and axial electric field show a compact pinch column and strong localized fields during the stagnation phase. The motion of the outermost current sheath agrees with LLNL fully kinetic results within about ten percent over the axial range where comparison is available, which gives confidence that the global current drive and inductive response are captured correctly.

Using the resolved plasma parameters, we have computed the deuterium-deuterium neutron yield directly from the simulated ion distribution. The total yield from the hybrid model is about $0.296 \times 10^{7}$, which lies within the same order of magnitude as the value reported for fully kinetic simulations of a similar device and remains several orders of magnitude larger than previous hybrid DPF calculations. Given the sensitivity of the yield to the assumed electron temperature closure, this level of agreement should be interpreted as an order-of-magnitude validation rather than as a precise prediction. Even with this caveat, the results demonstrate that the present model can approach fully kinetic performance for integral quantities such as neutron production while retaining a much lower computational cost.

The influence of additional electron physics has also been examined. Including the electron pressure gradient term slightly speeds up the rundown, advances the pinch time, and produces a broader yet well formed high density region. Adding the Hall term sharpens the sheath and strengthens local acceleration, but it also makes the late radial phase more sensitive to noise, which can trigger small scale structure and numerical instability. A predictor-corrector update for the current and a density based limiter on the pressure term keep the extended model stable, while the simpler configuration without the Hall contribution is already sufficient for quantitative comparison with reference data and for neutron yield estimation.

Overall, the results demonstrate that a fully electromagnetic ion PIC / electron fluid approach can provide a practical and predictive tool for dense plasma focus studies, bridging the gap between fast fluid models and expensive fully kinetic simulations. Future work will focus on improving the electron temperature treatment, refining the handling of Hall physics, extending the method to three dimensional geometries, and applying the framework to parameter scans and optimization studies for high yield DPF configurations.

%
% Each of the commands below will create an unnumbered section with the appropriate heading.
% Remove any sections that are not relevant for your article.
% All sections except suppdata will be removed if the [anonymous] option is used.
% See iopjournal-guidelines.pdf for more information.
%

\ack{
       The authors acknowledge the support from National Natural Science Foundation of China (Grant No.5247120164); National Natural Science Foundation of China (Grant No.12475256).
}

% \funding{}
% This section is a list of funder names and grant numbers

% \roles{Sample text inserted for demonstration.}
% List author names and the contributions made to the article, using terms from the NISO Contributor Roles Taxonomy (CRediT) https://credit.niso.org

\data{
       The data that support the findings of this study is available from the corresponding author upon reasonable request.
       }
% For more information on IOP Publishing's research data policy see: https://publishingsupport.iopscience.iop.org/questions/research-data/

% \suppdata{Sample text inserted for demonstration.}

% \section*{References}
\bibliographystyle{iopart-num}
\bibliography{References}

\end{document}